\documentclass[useAMS,usenatbib,usegraphicx]{mn2e}

\usepackage{upgreek}
\usepackage{amsmath}
\usepackage{amssymb}
\usepackage{txfonts}
\usepackage[colorlinks=true,citecolor=blue,linkcolor=blue,urlcolor=blue]{hyperref}
\newcommand{\dd}{\rmn{d}}

\renewcommand{\pi}{\upi}

\newcommand{\ext}{{\rmn{ext},i}}

\DeclareMathOperator{\erf}{erf}

\graphicspath{{./figures/}}

\begin{document} 

\title[Constraints from potential cluster function]{Constraints on $\bmath{\Omega_\mathrm{m}}$ and $\bmath{\sigma_8}$ from the potential-based cluster temperature function}

\author[C.~Angrick et al.]{Christian Angrick$^1$\thanks{E-mail: angrick@uni-heidelberg.de}, 
Francesco Pace$^2$, Matthias Bartelmann$^1$ and Mauro Roncarelli$^{3,4}$\\
$^1$Zentrum f\"ur Astronomie, Institut f\"ur Theoretische Astrophysik, Universit\"at Heidelberg, Philosophenweg~12, 
D-69120~Heidelberg, Germany\\
$^2$Jodrell Bank Centre for Astrophysics, School of Physics and Astronomy, The University of Manchester, Manchester, 
M13~9PL, UK\\
$^3$Dipartimento di Fisica e Astronomia, Universit\`a di Bologna, viale Berti Pichat 6/2, I-40127 Bologna, Italy\\
$^4$Istituto Nazionale di Astrofisica (INAF)-Osservatorio Astronomico di Bologna, via Ranzani 1, I-40127 Bologna, 
Italy}

\date{Accepted 2015 August 31. Received 2015 July 13; in original form 2015 April 13}

\maketitle

\begin{abstract}
The abundance of galaxy clusters is in principle a powerful tool to constrain cosmological parameters, especially $\Omega_\rmn{m}$ and $\sigma_8$, due to the exponential dependence in the high-mass regime. While the best observables are the X-ray temperature and luminosity, the abundance of galaxy clusters, however, is conventionally predicted as a function of mass. Hence, the intrinsic scatter and the uncertainties in the scaling relations between mass and either temperature or luminosity lower the reliability of galaxy clusters to constrain cosmological parameters. In this article, we further refine the X-ray temperature function for galaxy clusters by Angrick et al., which is based on the statistics of perturbations in the cosmic gravitational potential and proposed to replace the classical mass-based temperature function, by including a refined analytic merger model and compare the theoretical prediction to results from a cosmological hydrodynamical simulation. Although we find already a good agreement if we compare with a cluster temperature function based on the mass-weighted temperature, including a redshift-dependent scaling between mass-based and spectroscopic temperature yields even better agreement between theoretical model and numerical results. As a proof of concept, incorporating this additional scaling in our model, we constrain the cosmological parameters $\Omega_\rmn{m}$ and $\sigma_8$ from an X-ray sample of galaxy clusters and tentatively find agreement with the recent cosmic microwave background based results from the \textit{Planck} mission at 1$\sigma$-level.
\end{abstract}

\begin{keywords}
cosmology: theory -- methods: analytical -- dark matter -- cosmological parameters -- galaxies: clusters: general
\end{keywords}

\section{Introduction}
\label{sec:introduction}
Since the evolution of galaxy clusters is mainly driven by gravity and dark-matter (DM) physics, galaxy clusters provide in principle 
reliable information about cosmological parameters, especially on the density of matter in the Universe and the amplitude of initial density perturbations, expressed by $\Omega_\rmn{m}$ and $\sigma_8$, respectively. The usual procedure is 
to fit a theoretically motivated X-ray temperature function to X-ray samples obtained by analysing surveys conducted 
with X-ray telescopes like \textit{XMM Newton} and \textit{Chandra}.

The statistics of DM haloes, however, could conventionally only be analytically predicted based on masses, but measurements 
yield usually the X-ray temperature and luminosity. This implies that scaling relations between mass and other 
observables have to be calibrated relying on other cluster observables like e.g.\ weak-lensing shear, which reduces 
the power of the theoretical prediction by introducing additional sources of scatter. Additionally, the mass of a galaxy cluster is a global quantity and hence ill-defined since galaxy clusters have neither a fixed boundary with a regular shape nor a well-defined centre.

\citet{Angrick2009} developed an approach to derive the X-ray temperature function from the statistics of 
perturbations in the cosmic gravitational potential and the relation between potential depth and X-ray temperature. Using an extension of the 
ellipsoidal-collapse model of \citet{Bond1996} by \citet{Angrick2010}, \citet{Angrick2012} refined their X-ray 
temperature function by replacing spherical- with ellipsoidal-collapse dynamics and including an analytical merger 
model that takes the temporal temperature increase due to mergers statistically into account.

The goal of this paper is to further refine the merger model and analyse how the agreement between their analytical 
prediction and numerical realizations of the temperature function depends on the temperature definition used when 
setting up the halo catalogue inferred from a numerical simulation. We also want to analyse how constraints on 
$\Omega_\rmn{m}$ and $\sigma_8$ using the potential-based temperature function are affected if differences between 
the theoretically motivated X-ray temperature that enters via the virial theorem and temperatures that are actually 
inferred from measurements with X-ray telescopes are taken into account.

The structure of this paper is as follows: In Sect.~\ref{sec:theoBackground}, we summarize the main ingredients of the 
cluster temperature function built upon the statistics of the cosmic gravitational potential including the 
ellipsoidal-collapse model by \citet{Bond1996}, slightly extended by \citet{Angrick2010}, and the refined analytic 
merger model based on the model by \citet{Angrick2012}.

In Sect.~\ref{sec:simulation}, we give an overview of the fully hydrodynamical numerical simulation we base our 
analysis upon and introduce the temperature definitions we work with in the remainder of this work.

We contrast our theoretical model to results from this numerical simulation for different temperature definitions in 
Sect.~\ref{sec:models} and analyse how taking into account a relation between these differently defined temperatures 
improves the agreement between semi-analytic model and simulations.

In Sect.~\ref{sec:constraints}, we discuss how we can use these findings to constrain the cosmological parameters 
$\Omega_\rmn{m}$ and $\sigma_8$ from a fit to an X-ray sample by \citet{Vikhlinin2009} using the \emph{C} statistic 
by \citet{Cash1979}.

We present our constraints on both parameters in Sect.~\ref{sec:results} and quantify how a relation between the X-ray 
temperature from the model and the one inferred from measurements affects constraints on both parameters. 
Additionally, as a proof of concept, we compare to recent results from the \textit{Planck} mission.

Finally, we give a short summary in Sect.~\ref{sec:summary} and provide an outlook on how to proceed further with our 
potential-based temperature function.

\upshape

\section{Theoretical background}
\label{sec:theoBackground}

In this section, we will give short overviews of the main ingredients that are needed to construct the X-ray 
temperature function without reference to mass.

\subsection{The ellipsoidal-collapse model}
\label{subsec:ellCollapse}

Here, we will present the essential steps towards the ellipsoidal-collapse model used later in this work. For 
detailed calculations, refer to \citet{Bond1996} and \citet{Angrick2010}.

Let $a_i=R_i/R_\rmn{pk}$ with $1\leq i\leq 3$ be the dimensionless principal axes of a homogeneous ellipsoid, where 
the $R_i$ are the ellipsoid's physical semi-major axes, and $R_\rmn{pk}$ is the size of a spherical top-hat 
corresponding to a mass $M=(4\pi/3)\rho_\rmn{b}R_\rmn{pk}^3$ with the cosmological background density $\rho_\rmn{b}$.

Their evolution is described by the following three coupled differential equations,
\begin{equation}
 \label{eq:basicEvolutionA}
\frac{\dd^2a_i}{\dd a^2}+\left[\frac{1}{a}+\frac{E^{\prime}(a)}{E(a)}\right]\frac{\dd a_i}
{\dd a}+\left[\frac{3\Omega_\rmn{m}}
{2a^5 E^2(a)}C_i(a)-\frac{\Omega_\Lambda}{a^2 E^2(a)}\right]a_i=0,
\end{equation}
where $a$ is the scale factor of the Universe, $E(a)$ its expansion function, and $E^{\prime}(a)$ its derivative with 
respect to $a$, $\Omega_\rmn{m}$ and $\Omega_\Lambda$ are the dimensionless density parameters of matter and the cosmological 
constant, respectively, in units of the critical density today and $C_i\equiv(1+\delta)/3+b_i/2+\lambda_\ext$.

Here, $\delta=a^3/\left(a_1 a_2 a_3\right)-1$ is the density contrast of the evolving ellipsoid, $b_i$ is the $i$th 
component of the \emph{internal shear} given by
\begin{equation}
 \label{eq:defineIntShear}
b_i(a)=a_1(a)\,a_2(a)\,a_3(a)\int_0^\infty\frac{\dd\tau}{[a_i^2(a)+1]\prod_{k=1}^3[a_k^2(a)+1]^{1/2}}-\frac{2}{3},
\end{equation}
and the $\lambda_\ext$ denote the components of the \emph{external shear}, which we approximate by the so-called 
\emph{hybrid model},
\begin{equation}
 \label{eq:defineExtShear}
 \lambda_{\ext}(a)\equiv
 \begin{cases}
  \dfrac{5}{4}b_i(a) &\text{if } a\leq a_{\rmn{ta},i},\\[2mm]
  \dfrac{D_+(a)}{D_+(a_{\rmn{ta},i})}\lambda_{\ext}(a_{\rmn{ta},i}) &\text{else,}
 \end{cases}
\end{equation}
where $a_{\rmn{ta},i}$ is the scale factor of turn-around of the $i$th axis, $D_+$ is the linear growth factor of 
matter perturbations, and the $\lambda_i$ are the eigenvalues of the Zel'dovich deformation tensor 
\citep{Zeldovich1970}. Using the hybrid model, we take into account that the ellipsoid's evolution is initially 
tightly bound to its vicinity until each axis finally decouples at its turn-around.

The initial conditions of equation~\eqref{eq:basicEvolutionA} are given by the \emph{Zel'dovich approximation},
\begin{align}
 a_i(a_0)&=a_0[1-\lambda_i(a_0)], \label{eq:initAxes}\\
\left.\frac{\dd a_i}{\dd a}\right|_{a_0}&=1-\lambda_i(a_0)-\left.\frac{\dd\ln D_+}{\dd\ln a}\right|_{a_0}
\lambda_i(a_0)\approx 1-2\lambda_i(a_0), \label{eq:initAxesVel}
\end{align}
since $D_+(a_0)\approx a_0$ for a small initial scale factor $a_0$ chosen to be $a_0=2\times10^{-5}$. Note that in 
this ellipsoidal-collapse model, the reference frames of both the ellipsoid and the gravitational shear coincide.

The initial values of the $\lambda_i$ at $a_0$ are given by
\begin{equation}
\label{eq:lambdas}
\lambda_1=\frac{\delta}{3}+\frac{\sigma(M)}{\sqrt{10\pi}},\qquad\lambda_2=\frac{\delta}{3},
\qquad\lambda_3=\frac{\delta}{3}-\frac{\sigma(M)}{\sqrt{10\pi}},
\end{equation}
where $\sigma(M)$ is the square root of the matter power spectrum's variance filtered with a circular top-hat function 
$W_M$ on the scale $R(M)=[2GM/(\Omega_\rmn{m} H_0^2)]^{1/3}$ given by
\begin{equation}
\label{eq:variance}
\sigma^2(M)=\int_0^\infty\frac{\dd k}{2\pi^2}k^2P_\delta(k)\hat{W}_M^2(k).
\end{equation}
Here, $M$ is the halo's virial mass, $G$ is the gravitational constant, $H_0$ is the Hubble constant, $P_\delta(k)$ is the matter power 
spectrum and $\hat{W}_M(k)$ is the Fourier transform of $W_M$.

To derive equation~\eqref{eq:lambdas}, we implicitly used that the most probable values for the initial ellipticity 
and the prolaticity are the best choices to describe a statistical average of haloes with mass $M$.

According to equation~\eqref{eq:basicEvolutionA}, the evolution of the three axes $a_i$ continues until the smallest 
axis $a_1$ finally collapses first. At that point, the halo's density would formally be infinitely large, and the 
evolution of the other two axes could no longer be followed. But physically, the ellipsoid's collapse should be stopped 
before due to virialization.

A proper virialization condition can be derived for each axis from the tensor virial theorem. It reads
\begin{equation}
 \label{eq:virCondition}
\left(\frac{a_i^{\prime}}{a_i}\right)^2=\frac{1}{a^2 E^2(a)}
\left(\frac{3\Omega_\rmn{m}}{2a^3}C_i-\Omega_\Lambda\right)
\quad\text{and}\quad a_i^{\prime}<0.
\end{equation}
If the former expression is fulfilled for the $i$th axis, its collapse is stopped by hand, i.e.\ $a_i$ is set constant and $a_i'$ to zero for the further evolution. The ellipsoid is considered to be completely collapsed when the condition 
\eqref{eq:virCondition} is fulfilled for the largest axis $a_3$. This defines the virialization scale factor $a_\rmn{v}$.

The critical overdensity $\delta_\rmn{c}$ and the virial overdensity $\Delta_\rmn{v}$ with respect to the 
\emph{critical density} are given by
\begin{equation}
\label{eq:defineDeltas}
\delta_\rmn{c}=\frac{D_+(a_\rmn{v})}{D_+(a_0)}\sum_{i=1}^3\lambda_i\quad\text{and}\quad
\Delta_\rmn{v}=\frac{a_\rmn{v}^3\,\Omega_\rmn{m}(a_\rmn{v})}{a_1(a_\rmn{v})\,a_2(a_\rmn{v})\,a_3(a_\rmn{v})},
\end{equation}
respectively. Note that both $\delta_\rmn{c}$ and $\Delta_\rmn{v}$ are mass dependent in this model unlike
results from the spherical-collapse model!

\subsection{The X-ray temperature function without reference to mass}
\label{subsec:tempFunction}

In the following, we will shortly describe how to evaluate the cluster temperature function derived from the statistics 
of minima in a Gaussian random field and incorporating ellipsoidal-collapse dynamics. For a detailed derivation, see 
\citet{Angrick2009,Angrick2012}.

Based on the statistics of Gaussian random fields \citep{Bardeen1986}, the number density of minima in the cosmic 
gravitational potential as a function of the linear potential depth $\Phi_\rmn{l}$ and the potential's Laplacian 
$\Delta\Phi$ can be derived analytically and is given by
\begin{equation}
\label{eq:tildeN}
\tilde{n}(\Phi_\rmn{l},\Delta\Phi)=\frac{1}{240\pi^3\sigma_1^3\sqrt{15\gamma}}(F_1+F_2)
\exp\left[-\frac{\left(2\sigma_1^2\Delta\Phi+\sigma_2^2\Phi_\rmn{l}\right)\Phi_\rmn{l}}{2\gamma}\right]
\end{equation}
with
\begin{align}
\label{eq:F1}
F_1&=2\sigma_2\left(5\Delta\Phi^2-16\sigma_2^2\right)\exp\left[-\frac{\left(6\sigma_0^2\sigma_2^2-5\sigma_1^4\right)
\Delta\Phi^2}{2\sigma_2^2\gamma}\right] \nonumber \\ 
&+\sigma_2\left(155\Delta\Phi^2+32\sigma_2^2\right)
\exp\left[-\frac{\left(9\sigma_0^2\sigma_2^2-5\sigma_1^4\right)\Delta\Phi^2}{8\sigma_2^2\gamma}\right], \\
\label{eq:F2}
F_2&=5\sqrt{10\pi}\Delta\Phi\left(\Delta\Phi^2-3\sigma_2^2\right)\exp\left(-\frac{\sigma_0^2\Delta\Phi^2}{2\gamma}
\right) \nonumber \\
&\times\left[\erf\left(\frac{\sqrt{5}\Delta\Phi}{2\sqrt{2}\sigma_2}\right)+\erf\left(\frac{\sqrt{5}\Delta\Phi}
{\sqrt{2}\sigma_2}\right)\right].
\end{align}
Here, the $\sigma_j$ with $0\leq j\leq 2$ are the \emph{spectral moments} of the potential power spectrum $P_\Phi(k)$ 
defined as
\begin{equation}
\label{eq:specMoments}
\sigma_j^2=\int_{k_\rmn{min}}^\infty \frac{\dd k}{2\pi^2}k^{2+2j}P_\Phi(k) \hat{W}_{\Phi,\,\Delta\Phi}^2(k),
\end{equation}
and $\gamma\equiv\sigma_0\sigma_2-\sigma_1^2$. The window function in the former equation is given by
\begin{equation}
\label{eq:filter}
\hat{W}_{\Phi,\,\Delta\Phi}(k)=\frac{5\left[3\sin u-u\left(3+u^2\right)\cos u\right]}{2 u^5},
\end{equation}
where $u\equiv kR$ with $R=\sqrt{-2\Phi_\rmn{l}/\Delta\Phi}$, which is the Fourier transform of the functional form of 
a homogeneous sphere's gravitational potential.

The lower integration boundary $k_\rmn{min}$ in equation~\eqref{eq:specMoments} is chosen for each pair 
$(\Phi_\rmn{l},\Delta\Phi)$ such that the number density \eqref{eq:tildeN} is maximized. This effective high-pass filter removes large 
modes of the potential and also of its gradient so that the potential of an object is defined with respect to its 
local environment and small objects with $\nabla\Phi\neq 0$ are brought to rest and are therefore counted correctly. 
These objects correspond to minima in the non-linearly evolved potential which, however, are not present in the linearly 
evolved one.

The potential power spectrum $P_\Phi(k)$ is related to the matter power spectrum $P_\delta(k)$ by
\begin{equation}
\label{eq:relateSpectra}
P_\Phi(k)=\frac{9}{4}\frac{\Omega_\rmn{m}^2}{a^2}\frac{H_0^4}{k^4}P_\delta(k).
\end{equation}
To obtain the number density of minima in the gravitational potential that belong to collapsed structures, we have to 
integrate equation~\eqref{eq:tildeN} over $\Delta\Phi$ accordingly,
\begin{equation}
\label{eq:critLapEll}
n(\Phi_\rmn{l})=\int_0^\infty\dd\Delta\Phi\,\tilde{n}(\Phi_\rmn{l},\Delta\Phi)\,
\uptheta_\rmn{H}[\Delta\Phi-\Delta\Phi_\rmn{c}(\Phi_\rmn{l},\Delta\Phi)],
\end{equation}
where $\uptheta_\rmn{H}$ is Heaviside's step function and the critical Laplacian $\Delta\Phi_\rmn{c}$ is given by
\begin{equation}
\label{eq:critLaplacian}
\Delta\Phi_\rmn{c}(\Phi_\rmn{l},\Delta\Phi)=\frac{3}{2}H_0^2\Omega_\rmn{m}
\frac{\delta_\rmn{c}(\Phi_\rmn{l},\Delta\Phi)}{a}.
\end{equation}
Here, $\delta_\rmn{c}$ is dependent on both $\Phi_\rmn{l}$ and $\Delta\Phi$ through equations~\eqref{eq:lambdas} and 
\eqref{eq:variance} by setting again $R=\sqrt{-2\Phi_\rmn{l}/\Delta\Phi}$ instead of $R(M)$.

Since galaxy clusters are highly non-linear objects, we have to relate the linear potential, $\Phi_\rmn{l}$, to a non-
linear one, $\Phi_\rmn{nl}$. This is realized by comparing the potential in the centre of a homogeneous ellipsoid at 
the time of virialization to the linearly propagated potential. The result is
\begin{equation}
 \label{eq:phiLinNonlinEll}
\frac{\Phi_\rmn{nl}}{\Phi_\rmn{l}}=\frac{a_\rmn{v}}{2\delta_0}
\frac{D_+(a_0)}{D_+(a_\rmn{v})}\int_0^\infty\frac{\dd\tau}
{\sqrt{\left[a_1^2(a_\rmn{v})+\tau\right]\left[a_2^2(a_\rmn{v})+\tau\right]\left[a_3^2(a_\rmn{v})+\tau\right]}},
\end{equation}
where $\delta_0$ is the initial overdensity inside the collapsing halo chosen such that the last axis virializes at $a_\rmn{v}$. Note that the 
ratio $\Phi_\rmn{nl}/\Phi_\rmn{l}$ depends on both $\Phi_\rmn{l}$ and $\Delta\Phi$ through the axes $a_i$ since the 
ellipsoidal-collapse model is initialized with the scale $R=\sqrt{-2\Phi_\rmn{l}/\Delta\Phi}$. However, since we need 
to relate a single linear potential to a non-linear one, we marginalize over the dependence on $\Delta\Phi$ by
\begin{equation}
\label{eq:averagePhiLin}
\langle\Phi_\rmn{l}\rangle_{\Delta\Phi}(\Phi_\rmn{nl})=\frac{\int_0^\infty\dd\Delta\Phi\,\Phi_\rmn{l}\,
\tilde{n}(\Phi_\rmn{l},\Delta\Phi)\,\uptheta_\rmn{H}[\Delta\Phi-\Delta\Phi_\rmn{c}(\Phi_\rmn{l},\Delta\Phi)]}
{\int_0^\infty\dd\Delta\Phi\,
\tilde{n}(\Phi_\rmn{l},\Delta\Phi)\,\uptheta_\rmn{H}[\Delta\Phi-\Delta\Phi_\rmn{c}(\Phi_\rmn{l},\Delta\Phi)]},
\end{equation}
where $\Phi_\rmn{l}$ is a function of $\Phi_\rmn{nl}$ and $\Delta\Phi$ via equation~\eqref{eq:phiLinNonlinEll}.

The non-linear potential depth can be related to an X-ray temperature $T$ via the virial theorem if we average over 
sufficiently many particle orbits. For particles near the centre, this yields
\begin{equation}
\label{eq:virTheorem}
k_\rmn{B}T=-\frac{1}{3}\mu m_\rmn{p}\Phi_\rmn{nl},
\end{equation}
where $k_\rmn{B}$ is Boltzmann's constant, $m_\rmn{p}$ is the proton mass and $\mu$ represents the mean molecular weight and is defined as
\begin{equation}
 \mu=\left(2X_\rmn{p}+\frac{3}{4}Y_\rmn{p}\right)^{-1},
\end{equation}
where $X_\rmn{p}=0.76$ and $Y_\rmn{p}=1-X_\rmn{p}$ are the primordial hydrogen and helium mass fractions, respectively, so that $\mu\approx 0.59$.

Summarising the above relations, the number density of galaxy clusters as a function of their X-ray temperature is 
given by
\begin{equation}
 \label{eq:numDensT}
n_\rmn{vir}(T)\,\dd T=\underbrace{n(T\stackrel{\eqref{eq:virTheorem}}{\rightarrow}\Phi_\rmn{nl}
\stackrel{\eqref{eq:averagePhiLin}}{\rightarrow}\langle\Phi_\rmn{l}\rangle_{\Delta\Phi})}_{\eqref{eq:critLapEll}}
\left|\frac{\dd\langle\Phi_\rmn{l}\rangle_{\Delta\Phi}}
{\dd\Phi_\rmn{nl}}\frac{\dd \Phi_\rmn{nl}}{\dd T}\right|\,\dd T,
\end{equation}
where the derivative $\dd\langle\Phi_\rmn{l}\rangle_{\Delta\Phi}/\dd\Phi_\rmn{nl}$ has to be calculated numerically 
from equations~\eqref{eq:phiLinNonlinEll} and \eqref{eq:averagePhiLin}.

\subsection{The influence of mergers on the temperature function}
\label{subsec:mergers}

So far, we have only taken into account virialized structures for which the relation \eqref{eq:virTheorem} holds. 
However, due to hierarchical structure formation, smaller DM haloes merge to form larger haloes. Since these 
mergers induce a rise in the X-ray temperature function \citep*[see e.g.][]{Randall2002}, we will include this effect 
statistically in our framework by a simple parameter-free merger model. The following model slightly extends the model 
by \citet{Angrick2012} to account for the change of the cluster temperature function with redshift.

Starting from the number density of virialized galaxy clusters at a given temperature and redshift, $n_\mathrm{vir}(T,z)$, we calculate two correction terms. First, we add the clusters that reach a temperature $T$ only due to a temperature boost $\Delta T$ induced by a merger, and second, we subtract those that would have a temperature $T$ if they were virialized, but have a temperature higher than $T$ due to a merger.

Denoting the first contribution by $n_+(T)$ and the second by $n_-(T)$, the final halo population $n(T,z)$ can be modelled as
\begin{equation}
\label{eq:numDensMerger}
n(T,z)=n_\rmn{vir}(T,z)+n_+(T,z)-n_-(T,z),
\end{equation}
where $n_+(T,z)$ is given by
\begin{align}
\nonumber
n_+(T,z)=&\,\int_z^{z+\Delta z}\dd z'\int_0^\infty \dd\Delta M\int_0^\infty\dd M\, n[T_\rmn{vir}(M,z'),z']\\
\nonumber
&\times p(M,\Delta M,z')\,\updelta_\rmn{D}[T-T_\rmn{vir}(M,z')-\Delta T(M,\Delta M,z')]\\
\label{eq:nPlus}
&\times\uptheta_\rmn{H}(M-\Delta M)
\end{align}
and $n_-(T,z)$ by
\begin{equation}
\label{eq:nMinus}
n_-(T,z)=\int_z^{z+\Delta z}\dd z'\int_0^{M(T,z')}\dd\Delta M\, n(T,z')\,p[M(T,z'),\Delta M, z'].
\end{equation}
In the two former equations, there are various quantities that have to be defined in the following.

The merger rate $p(M,\Delta M,z)\,\dd\Delta M\,\dd z$ yields the number of mergers of haloes with mass $M$ with other 
haloes in the mass range $[\Delta M,\Delta M+\dd\Delta M]$ and in the redshift range $[z,z+\dd z]$ and is given by 
\begin{align}
\nonumber
p(M,\Delta M, z)=&\,\frac{1}{\sqrt{2\pi}}\left[\frac{S_1}{S_2(S_1-S_2)}\right]^{3/2}
\exp\left[-\frac{\omega^2(S_1-S_2)}{2S_1S_2}\right]\\
\label{eq:mergerRate}
&\times\left|\frac{\dd S_2}{\dd\Delta M}\frac{\dd\omega}{\dd z}\right|
\end{align}
\citep{Lacey1993}, where $S_1\equiv\sigma^2(M)$, $S_2\equiv\sigma^2(M+\Delta M)$ (cf.\ equation~\ref{eq:variance}), and 
$\omega\equiv\delta_\rmn{c}(z)/D_+(z)$. Here, $\delta_\rmn{c}(z)$ is the critical overdensity of the \emph{spherical}-collapse
model. Heaviside's $\uptheta$-function in equation~\eqref{eq:nPlus} ensures that $M\geq\Delta M$ so that mergers 
between two objects with masses $M$ and $\Delta M$ are only counted once by demanding that the mass ratio between main and merging object is always larger than unity.

The redshift interval $\Delta z$ is connected to the sound-crossing time $t_\rmn{sc}$ by
\begin{equation}
\label{eq:soundCrossing}
t_\rmn{sc}\equiv\frac{R}{c_\rmn{s}}=\frac{1}{H_0}\int\limits_z^{z+\Delta z}\frac{\dd z'}{E(z')\,(1+z')},
\end{equation}
where
\begin{equation}
\label{eq:radiusMass}
R(M)=\left(\frac{3M a_1a_2a_3}{4\pi\rho_\rmn{b}}\right)^{1/3}\quad\text{and}\quad c_\rmn{s}=\sqrt{\frac{5}{3}
\frac{k_\rmn{B}T}{\mu m_\rmn{p}}},
\end{equation}
are the cluster's mean radius and the sound speed, respectively.

\begin{figure*}
\centering
\includegraphics[width=0.33\textwidth]{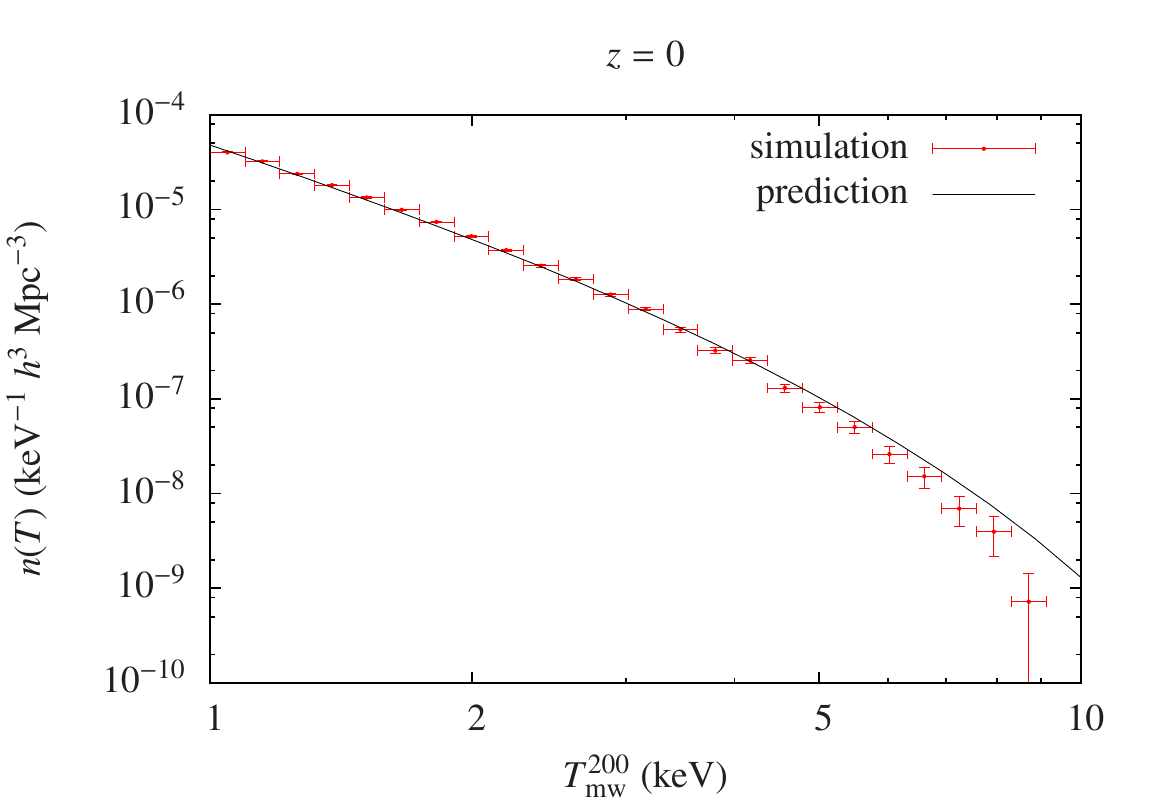}~\includegraphics[width=0.33\textwidth]{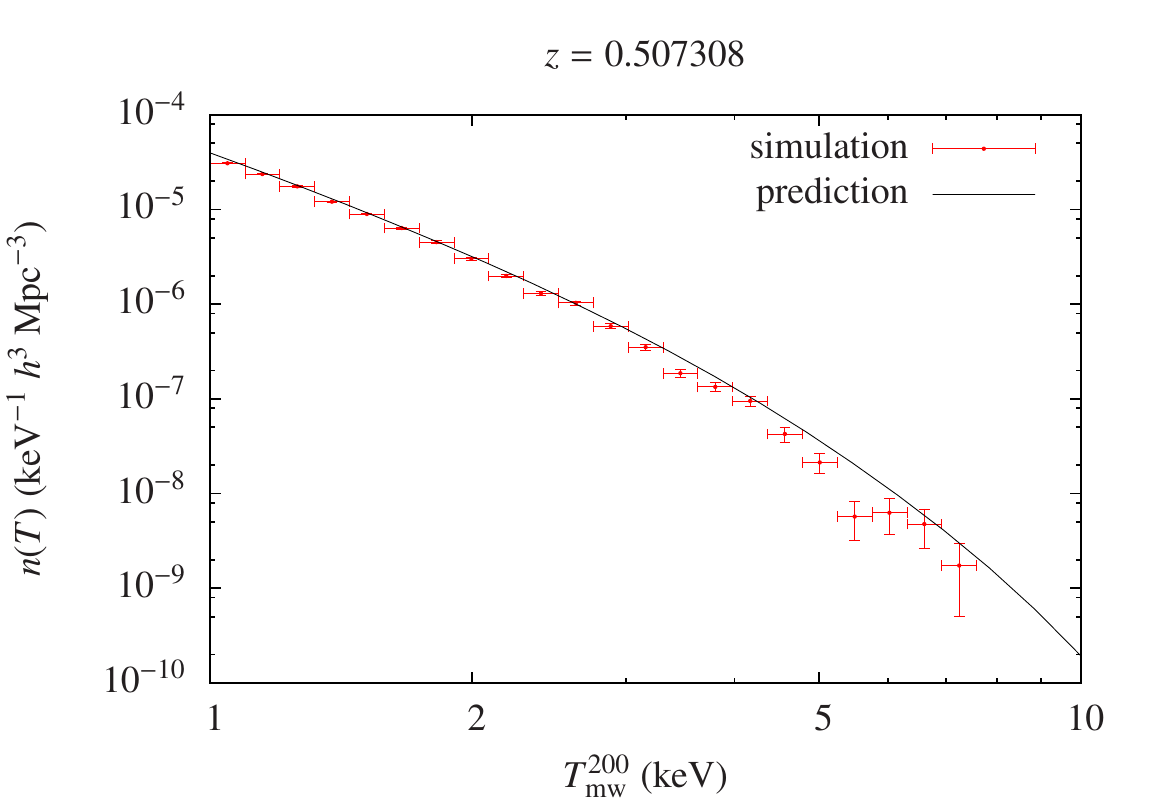}~
\includegraphics[width=0.33\textwidth]{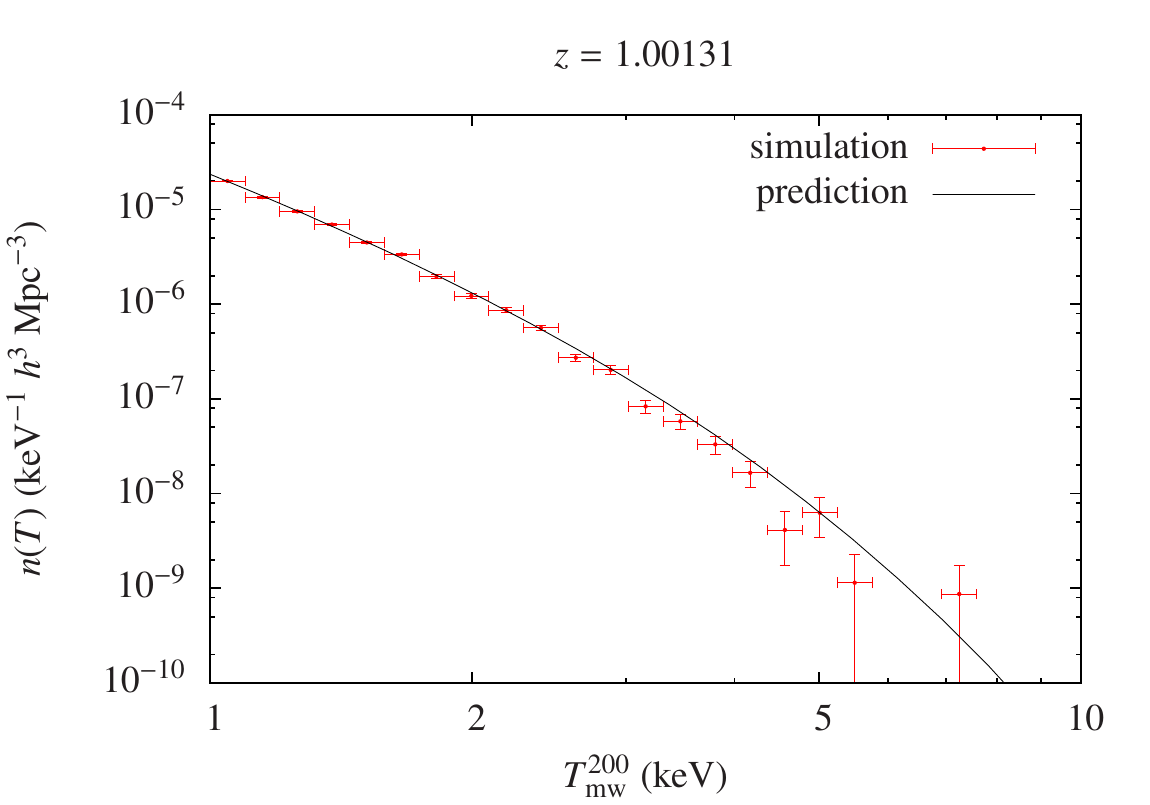}\\[2mm]
\includegraphics[width=0.33\textwidth]{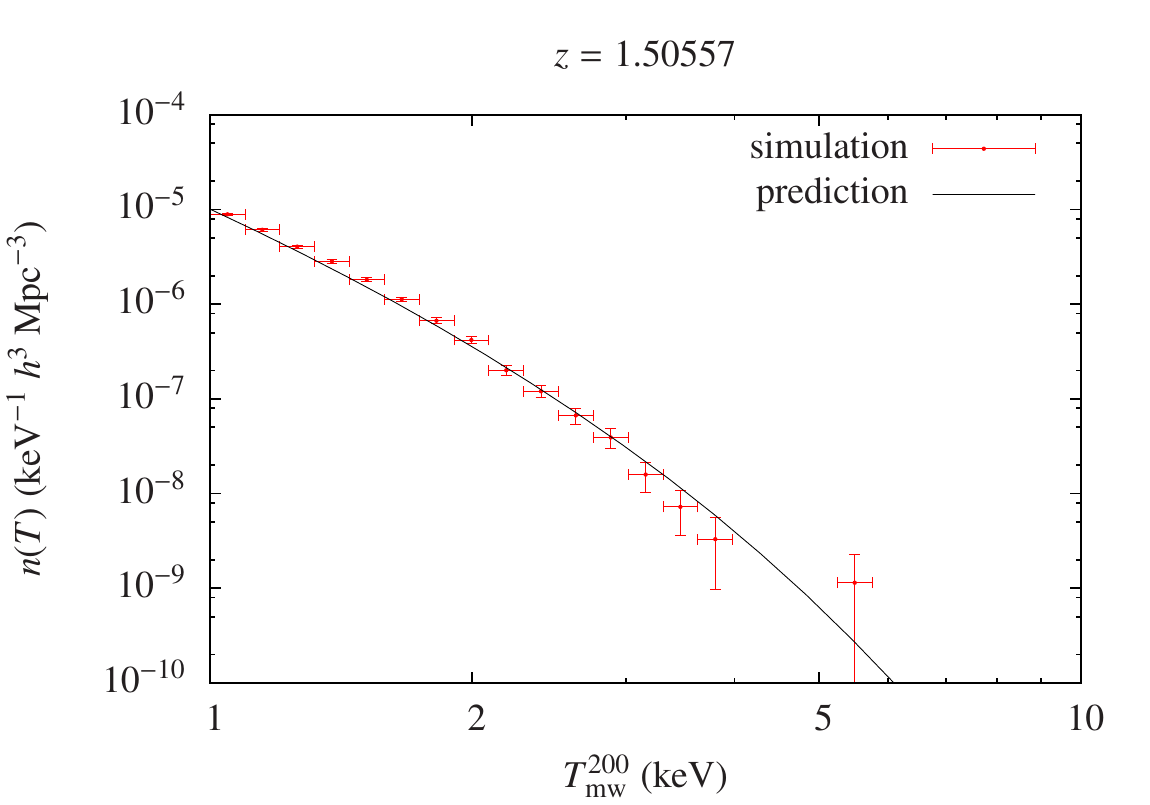}~\includegraphics[width=0.33\textwidth]{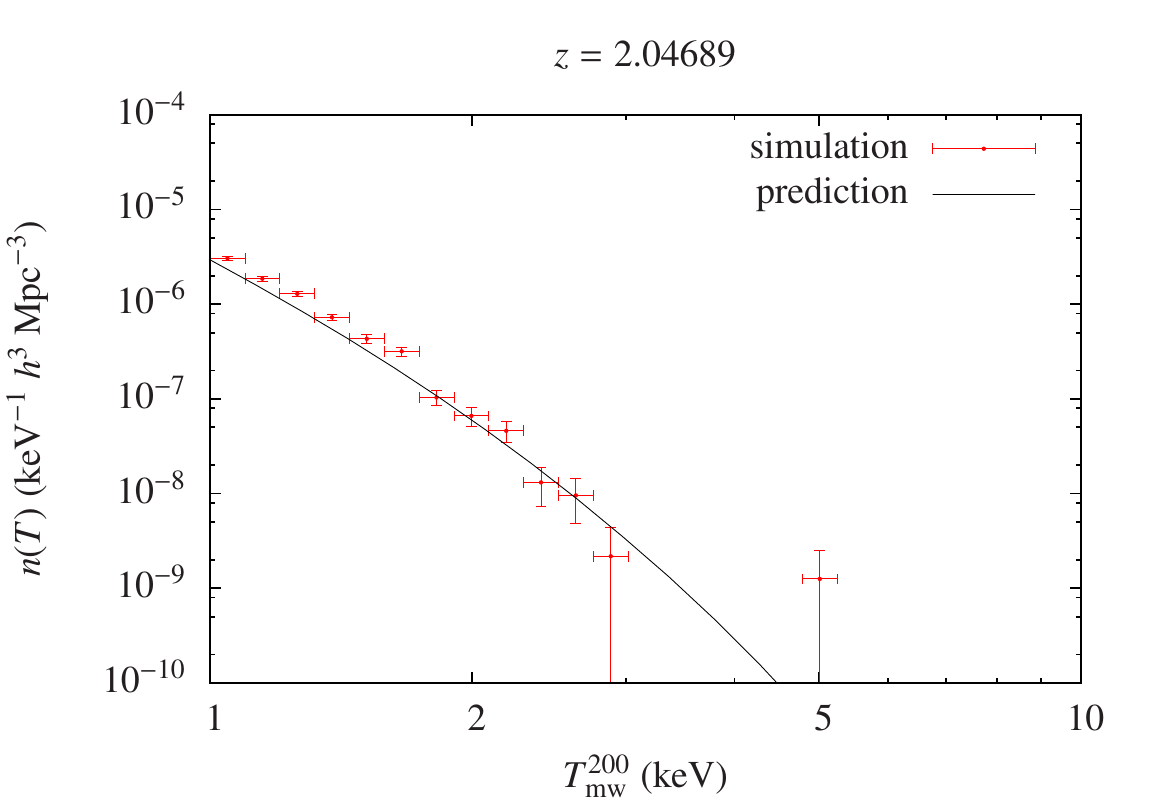}
\caption{The analytic temperature function $n(T)$ including merger effects (equation~\ref{eq:numDensMerger}) and the 
temperature function inferred from the numerical simulation based on $T_\rmn{mw}^{200}$ at five different 
redshifts.}
\label{fig:Tmw200}
\end{figure*}

The theoretical temperature-mass relation for virialized ellipsoidal haloes is given by
\begin{equation}
\label{eq:massTempRel}
k_\rmn{B} T_\rmn{vir}(M,z)=\mu m_\rmn{p}\left(\frac{\Omega_\rmn{m}H_0^2 G^2 M^2}{128}\right)^{1/3}\int_0^\infty
\frac{\dd\tau}{\prod_{k=1}^3\left[a_k^2(M,z)+\tau\right]^{1/2}}.
\end{equation}
Dirac's delta-distribution $\updelta_\rmn{D}$ in equation~\eqref{eq:nPlus} ensures that only those clusters contribute 
to the integral whose temperature $T$ is the sum of their temperature based on virial equilibrium, $T_\rmn{vir}$, and 
a temperature increase $\Delta T$ due to the merger given by
\begin{equation}
\label{eq:tempIncrease}
k_\rmn{B}\,\Delta T(M,\Delta M,z)=-\frac{2}{3}\frac{\mu m_\rmn{p}\,\Delta M\,\langle\Phi\rangle(M,z)}{M},
\end{equation}
where $\langle\Phi\rangle(M,z)$ is the mean potential on the surface of an ellipsoid with mass $M$, virialized at 
redshift $z$,
\begin{align}
\nonumber
\langle\Phi\rangle&(M,z)=-\frac{3}{4}\frac{MG}{R_\rmn{pk}(M)}\left\{\int_0^\infty
\frac{\dd\tau}{\prod_{k=1}^3\left[a_k^2(M,z)+\tau\right]^{1/2}}\right.\\
\label{eq:phiAverage}
&\left.-\frac{1}{3}\sum_{j=1}^3a_j^2(M,z)\int_0^\infty\frac{\dd\tau}{\left[a_j^2(M,z)+\tau\right]\prod_{k=1}^3
\left[a_k^2(M,z)+\tau\right]^{1/2}}\right\}.
\end{align}
$n_-(T)$ can be calculated accordingly from equation~\eqref{eq:nMinus} using equation~\eqref{eq:mergerRate} for 
$p(M,\Delta M,z)$, equation~\eqref{eq:soundCrossing} for $\Delta z$, and inverting equation~\eqref{eq:massTempRel} 
to arrive at $M(T,z)$. Note that in equation~\eqref{eq:nMinus}, $T$ is simply the cluster's virial temperature $T_\rmn{vir}$.

\section{The simulation}
\label{sec:simulation}

Hydrodynamical simulations are important numerical tools to describe in detail the evolution of cosmic structures in 
the Universe taking into account gas physics. 
In this section we describe in detail the numerical set-up adopted in this work. In particular, we will describe the 
simulations used, how haloes were extracted and how we built the halo catalogues with their X-ray properties.

The hydrodynamical simulation used in this work is part of a bigger 
set of simulations (BINGS, \textit{B}aryons \textit{i}n \textit{n}on-\textit{G}aussian \textit{s}imulations) originally proposed to study the effect of baryons in cosmologies with non-Gaussian initial conditions. They 
represent an extension of the DM only simulations analysed to study several aspects of the large-scale 
structures of this kind of cosmologies \citep{Grossi2009}. In addition, the DM-only simulations with the same 
cosmological parameters and box size have been already employed for studies on the SZ effect and X-ray emission 
\citep{Roncarelli2010}. For this work, we use exclusively the simulation run with Gaussian initial conditions, 
also in the light of recent observations of the cosmic microwave background (CMB) by the \textit{Planck} mission 
\citep{Planck2015}.

The simulations are based on the parallel cosmological TreePM-SPH code Gadget-2 \citep{Springel2005} using an 
entropy-conserving formulation of smoothed-particle hydrodynamics (SPH; \citealp{Springel2002}). The simulation run includes radiative cooling, star 
formation with associated feedback processes \citep{Springel2003} and heating by a uniform, time-dependent 
ultraviolet background. The simulation also takes into account a multiphase model for star-formation and feedback 
processes due to supernovae-driven galactic winds \citep{Springel2003,DiMatteo2008}.

The numerical set-up assumes a box of 1.2 $h^{-1}$ Gpc comoving with $2\times960^3$ DM and gas particles. The underlying
cosmological model is a standard $\Lambda$ cold dark matter model with the cosmological parameters derived by the 5 yr 
\textit{WMAP} results. We report them in Table~\ref{tab:params}. DM particles have a mass of $1.171\times10^{11}\ h^{-1}\ \text{M}_\odot$, 
and gas particles have a mass of 
$2.38\times10^{10}\ h^{-1}\ \text{M}_\odot$ (at the redshift of the initial conditions, $z_\rmn{ini}=60$). The gravitational force is computed assuming a Plummer-equivalent softening 
length of $\epsilon=25\ h^{-1}$ kpc. The evolution of particles is followed till the present time ($z=0$). To create the 
initial conditions of our simulation, the transfer function by \citet{Eisenstein1998} was used.

\begin{table}
\caption{Set of cosmological parameters at $z=0$ employed in the simulations. In the lower part 
additional characteristics of the simulations are reported.}
\label{tab:params}
\begin{center}
\begin{tabular}{lcc}
\hline
Parameter       & Symbol & Value \\ \hline
Hubble parameter (100 km s$^{-1}$ Mpc$^{-1}$) & $h$ & 0.72 \\
Amplitude of fluctuations at 8 $h^{-1}$ Mpc & $\sigma_8$         & 0.8 \\
Baryon density                            & $\Omega_\rmn{b}$   & 0.044 \\
Total matter density                            & $\Omega_\rmn{m}$   & 0.26 \\
Dark energy density                       & $\Omega_\Lambda$ & 0.74 \\
 \hline
Initial redshift                           & $z_\rmn{ini}$      & 60 \\
Cubic box length ($h^{-1}$ Mpc)            & $L_\rmn{box}$      & 1200 \\
Total number of particles                  & $N_\rmn{part}$     & $2\times 960^3$ \\
Mass of DM particles ($h^{-1}\ \text{M}_\odot$)  & $m_\rmn{dm}$       & $1.171\times10^{11}$ \\
Mass of gas particles at $z_\rmn{ini}$ ($h^{-1}\ \text{M}_\odot$) & $m_\rmn{gas}$      & $2.38\times10^{10}$ \\ \hline
\end{tabular}
\end{center}
\end{table}

Haloes were initially identified with a Friend-of-Friend algorithm \citep{Davis1985} with linking length $b=0.2$ 
times the mean interparticle distance and in a second step, using the SUBFIND algorithm 
\citep{Springel2001,Dolag2009}, we evaluated spherical-overdensity masses for each halo centred at the deepest 
potential point. We computed two different overdensities, $\Delta=200$ and $\Delta=500$ times the critical density. The corresponding radii are $R_{200}$ and $R_{500}$, respectively.
We extracted our catalogues at different redshifts, namely $z=0$, $z\approx 0.5$, $z\approx 1$, $z\approx 1.5$, and 
$z\approx 2$, and considered only objects with at least 100 DM particles to incorporate only haloes in the further analysis whose X-ray profiles are resolved well enough..

To evaluate the X-ray properties of each halo, we took into account all its gas particles, computed the relevant 
quantities for each individual particle and summed them up to have the integrated value for the object considered. For 
what concerns us, we will limit our discussion to the evaluation of the halo temperature. Starting from the internal energy $u$ of a given SPH particle, we evaluate its temperature with the following relation,
\begin{equation}
 T_i=\frac{2}{3k_\rmn{B}}\mu m_\rmn{p}u_i.
\end{equation}

For a given object we can define three different temperatures: the \emph{mass-weighted} one, $T_\rmn{mw}$, 
\citep{Kang1994,Bartelmann1996,Mathiesen2001}, the \emph{emission-weighted} one, $T_\rmn{ew}$, 
\citep{Kang1994,Mathiesen2001}, and the \emph{spectroscopic-like} temperature $T_\rmn{sl}$ \citep{Mazzotta2004}. 
They are defined as
\begin{align}
\label{eq:Tmw}
T_\rmn{mw}&=\frac{\int\dd V\,\rho T}{\int\dd V\,\rho},\\
\label{eq:Tew}
T_\rmn{ew}&=\frac{\int\dd V\,\rho^{2}T}{\int\dd V\,\rho^{2}},\\
\label{eq:Tsl}
T_\rmn{sl}&=\frac{\int\dd V\,\rho^{2}T^{1/4}}{\int\dd V\,\rho^{2}T^{-3/4}},
\end{align}
respectively, where $\rho$ is the density and $V$ is the integration volume (in this case the volume of the halo). In the following, we will consider a spherical integration volume with either $R_{200}$ or $R_{500}$ as a radius and indicate the corresponding temperature as, e.g., $T^{200}_\rmn{mw}$ or $T^{500}_\rmn{mw}$, respectively.

\begin{figure*}
\centering
\includegraphics[width=0.33\textwidth]{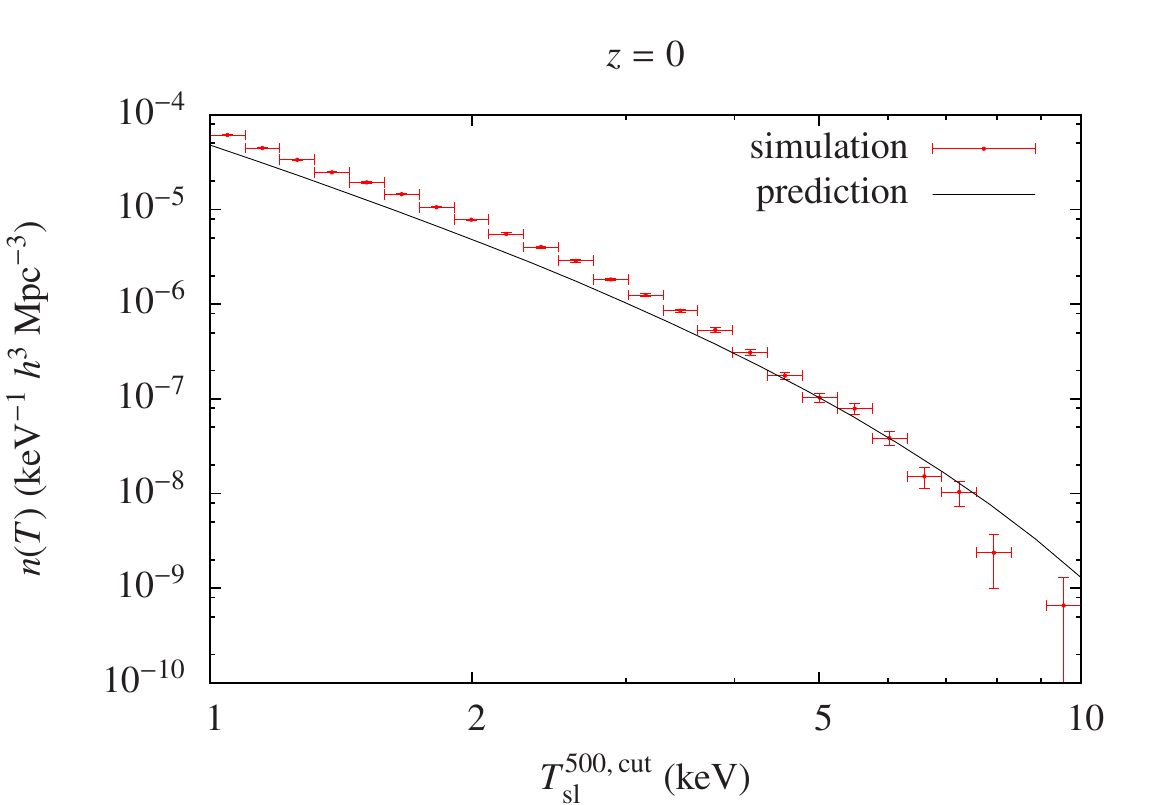}~\includegraphics[width=0.33\textwidth]{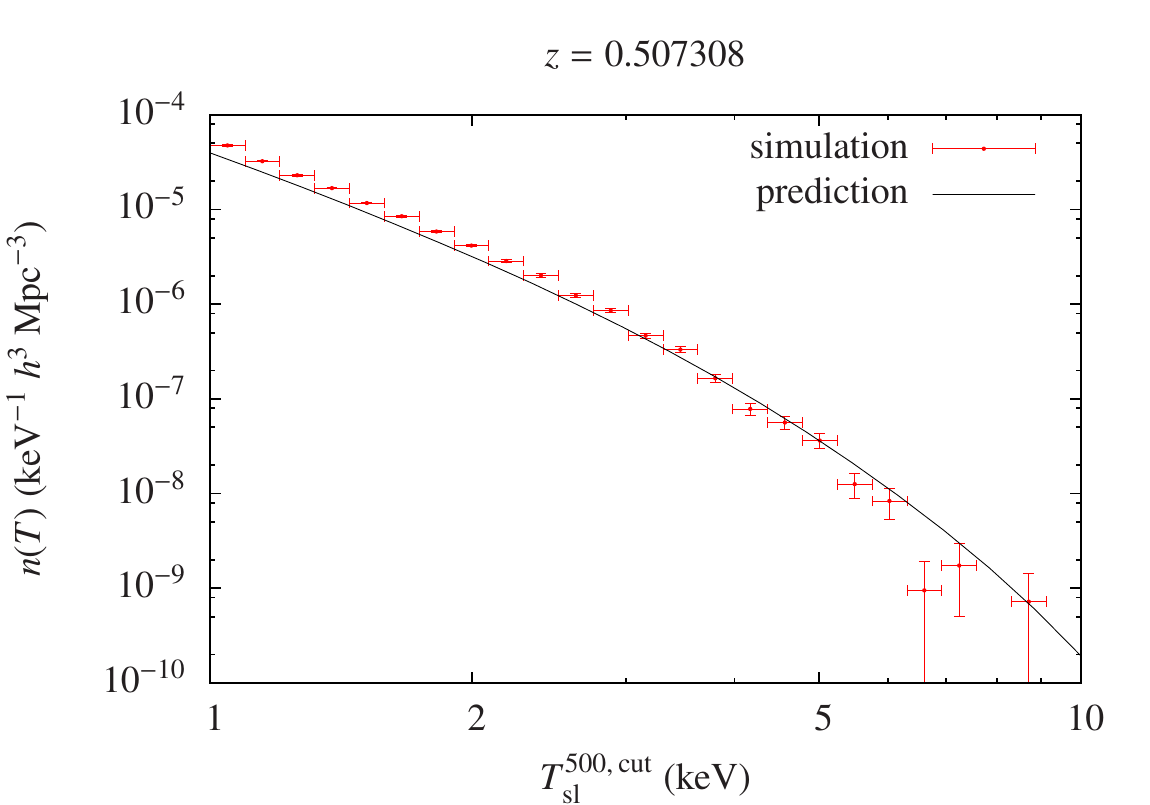}~
\includegraphics[width=0.33\textwidth]{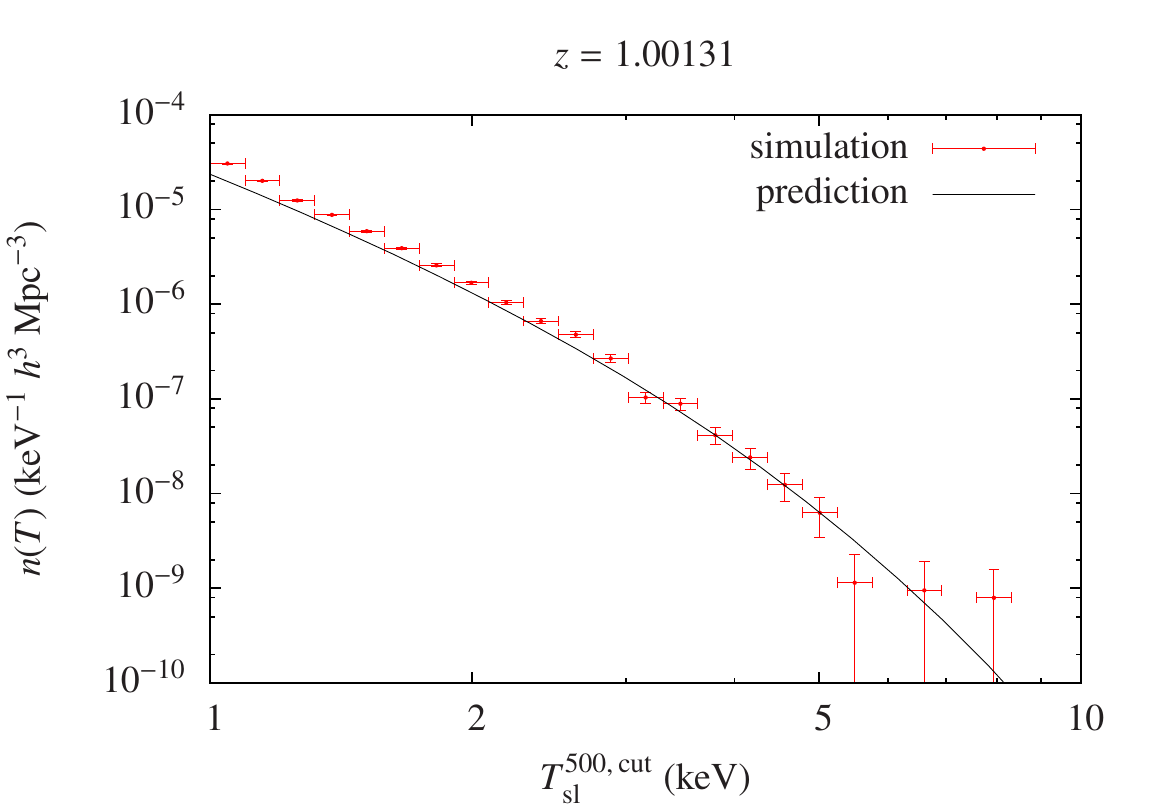}\\[2mm]
\includegraphics[width=0.33\textwidth]{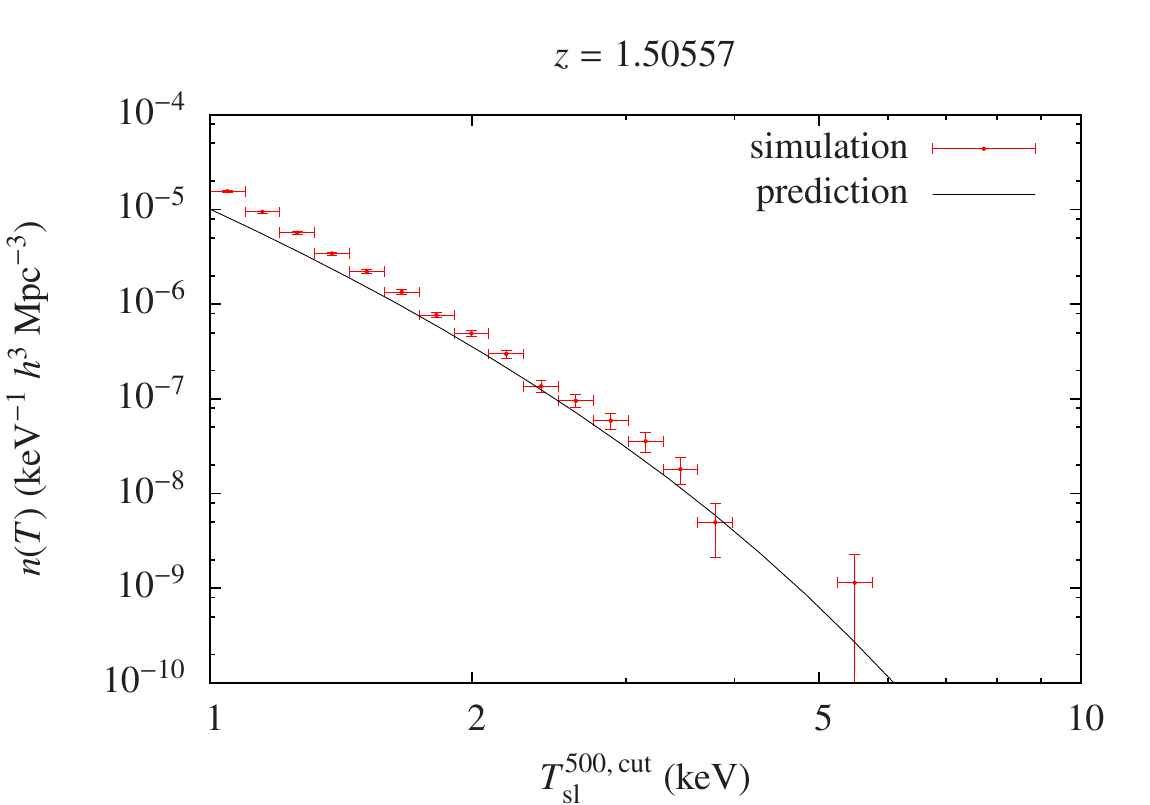}~\includegraphics[width=0.33\textwidth]{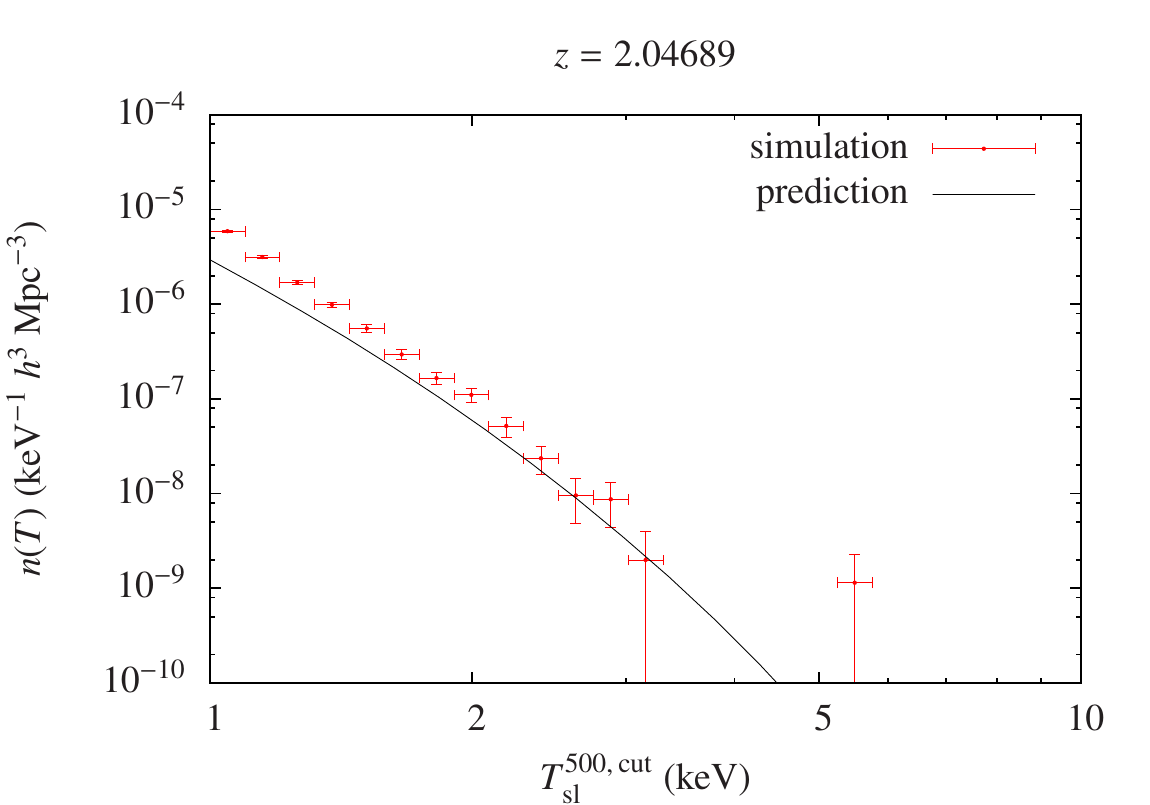}
\caption{Same as Fig.~\ref{fig:Tmw200}, but the temperature function inferred from the simulation is based on 
$T_\rmn{sl}^\rmn{500,\,cut}$.}
\label{fig:Tsl500cut}
\end{figure*}

Since the halo density is not simulated as a continuous quantity, we can replace the integration with a sum over all particles such that the operative definitions of the three different temperatures become
\begin{align}
T_\rmn{mw}&=\frac{\sum_i m_i T_i}{\sum_i m_i},\\
T_\rmn{ew}&=\frac{\sum_i m_i \rho_i T_i}{\sum_i m_i \rho_i},\\
T_\rmn{sl}&=\frac{\sum_i m_i \rho_i T_i^{1/4}}{\sum_i m_i \rho_i T_i^{-3/4}},
\end{align}
where $m_i$ and $\rho_i$ are the mass of the $i$th particle and the SPH estimate of the density at its position, respectively.\footnote{Note that there are more complicated definitions of the emission-weighted temperature, involving in its definition the 
cooling function \citep{Bryan1998,Frenk1999,Muanwong2001,Borgani2004}. However, since we will not use the emission-weighted temperature in the further discussion, we limit ourselves to the more simplified definition \eqref{eq:Tew} here.}

The mass-weighted temperature (equation~\ref{eq:Tmw}) has a relevant physical meaning since it is proportional to the total thermal energy of the cluster. However, it may differ significantly from what an observer would measure via an X-ray spectral analysis since the bremsstrahlung emissivity is proportional to the square of the gas density: the emission-weighted temperature (equation~\ref{eq:Tew}) takes this bias into account. The spectroscopic-like temperature (equation~\ref{eq:Tsl}), introduced by \citet{Mazzotta2004}, besides the emissivity bias, accounts also for the gas temperature gradient along the line of sight together with the shape of the energy response function of \textit{Chandra} and \textit{XMM-Newton} X-ray detectors.\footnote{When calculating $T_\rmn{ew}$ and $T_\rmn{sl}$, it is necessary to remove cold SPH particles ($T_i<0.5$~keV) from the computation to exclude the contribution of the cold dense phase of the ICM that has negligible emissivity.}

\section{Analytical model versus simulation}
\label{sec:models}

In the following, we want to compare the results of our analytical model from Sect.~\ref{sec:theoBackground} with 
results from the simulation presented in the previous section based on the two temperature definitions 
$T_\rmn{mw}^{200}$ and $T_\rmn{sl}^\rmn{500,\,cut}$ in the analysis, where `cut' in the superscript refers to the fact that the volume corresponding to the inner 15 per cent of the sphere's radius is cut out from the integration to obtain $T_\rmn{sl}^{500}$. The reason for cutting out the inner part will become clear shortly.

In Fig.~\ref{fig:Tmw200}, we show the X-ray temperature function based on $T_\rmn{mw}^{200}$ together with our semi-analytical prediction for five different redshifts. The theoretical curve matches the numerical results from the simulation very well for lower 
temperatures. At temperatures $\gtrsim4-5$~keV, however, the theoretical prediction lies above the temperature 
function from the simulation. A reason for the latter might be that haloes with a high X-ray temperature and therefore also a large mass are not well resolved 
in the simulation due to the limited number of large-scale modes in the simulation.

Since the X-ray temperature in our model is derived from the thermal energy of the virialized cluster and the mass-weighted temperature is proportional to the cluster's thermal energy, $T_\rmn{mw}^{200}$ is the definition used in the simulation that matches best the one in our model and hence is the appropriate quantity in this case.

In Fig.~\ref{fig:Tsl500cut}, we show the X-ray temperature function based on $T_\rmn{sl}^\rmn{500,\,cut}$ as inferred 
from the simulation together with our semi-analytic prediction since X-ray observers often fit the spectrum of a galaxy cluster in the interval $[0.15,1]R_{500}$. The theoretical prediction underestimates the number density of haloes at temperatures $\lesssim4-5$~keV, whereas for higher temperatures the theoretical curve lies above the results from the simulation at low redshifts. This might be another hint at missing high-mass objects in the simulation and incomplete statistics. Overall, the theoretical curve seems too flat compared to the numerical one inferred from the simulation.

\begin{figure*}
\centering
\includegraphics[width=0.33\textwidth]{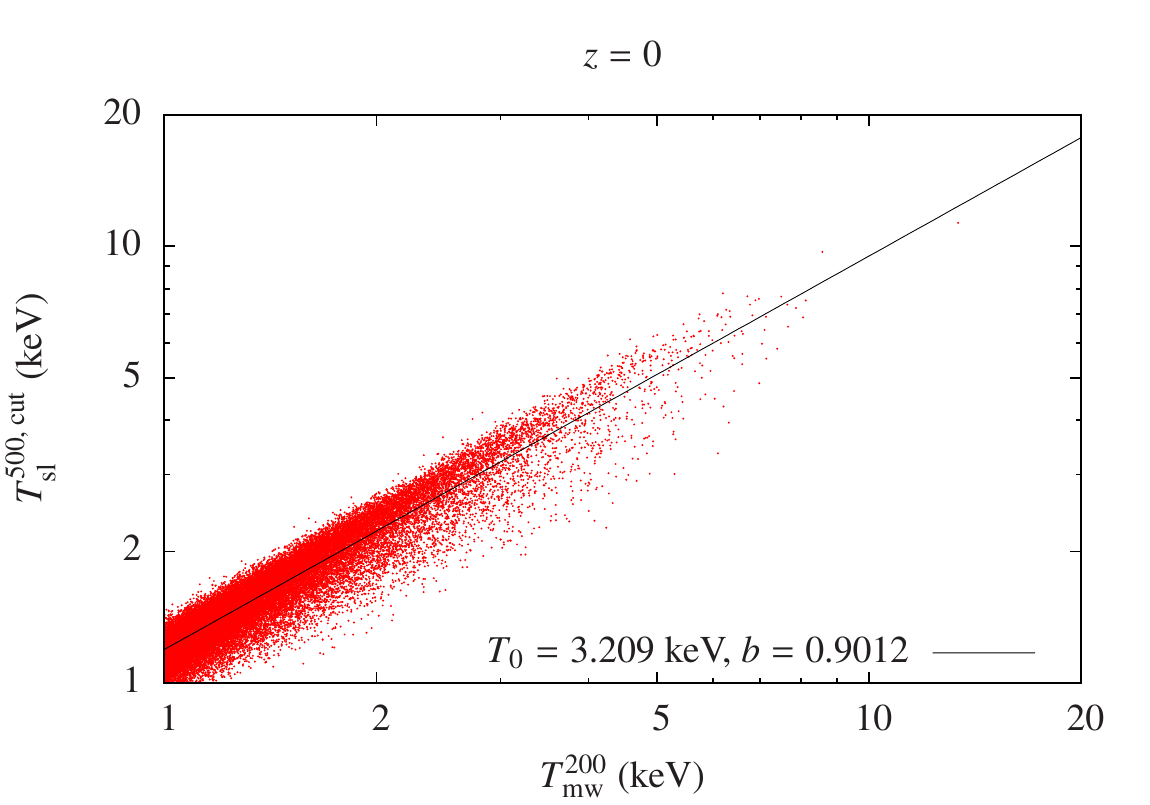}~\includegraphics[width=0.33\textwidth]{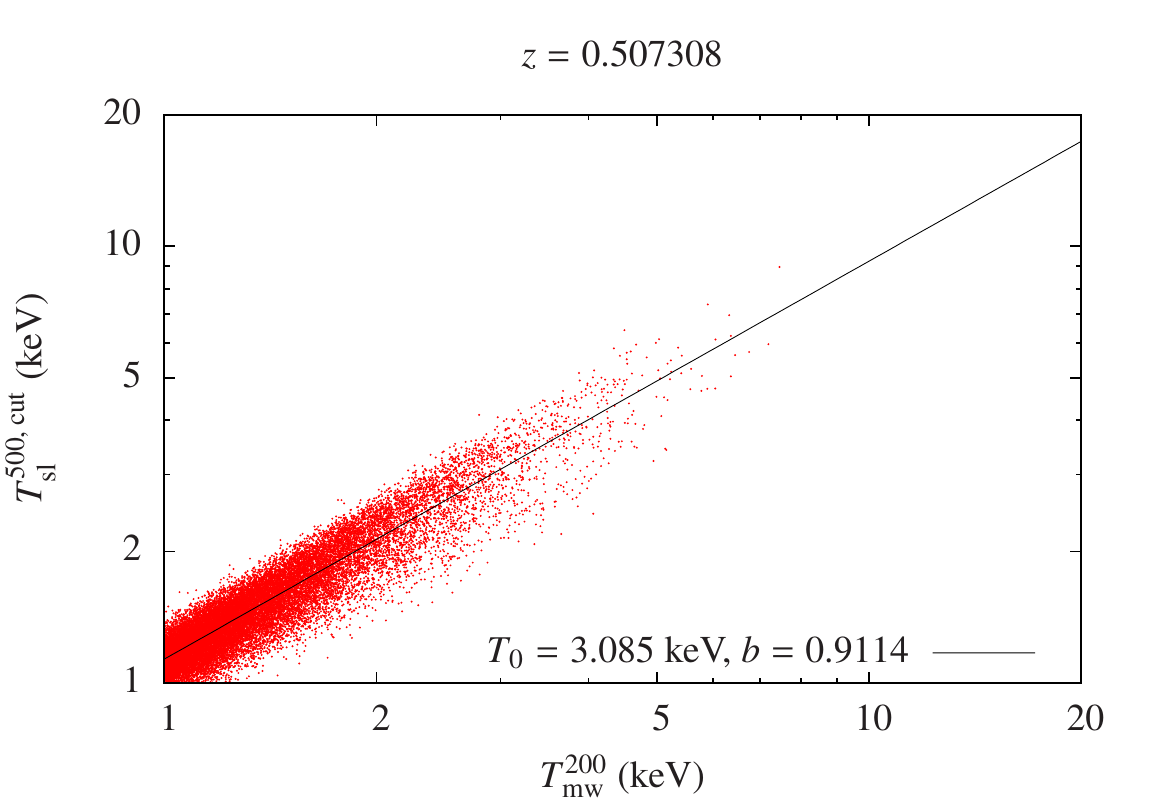}~
\includegraphics[width=0.33\textwidth]{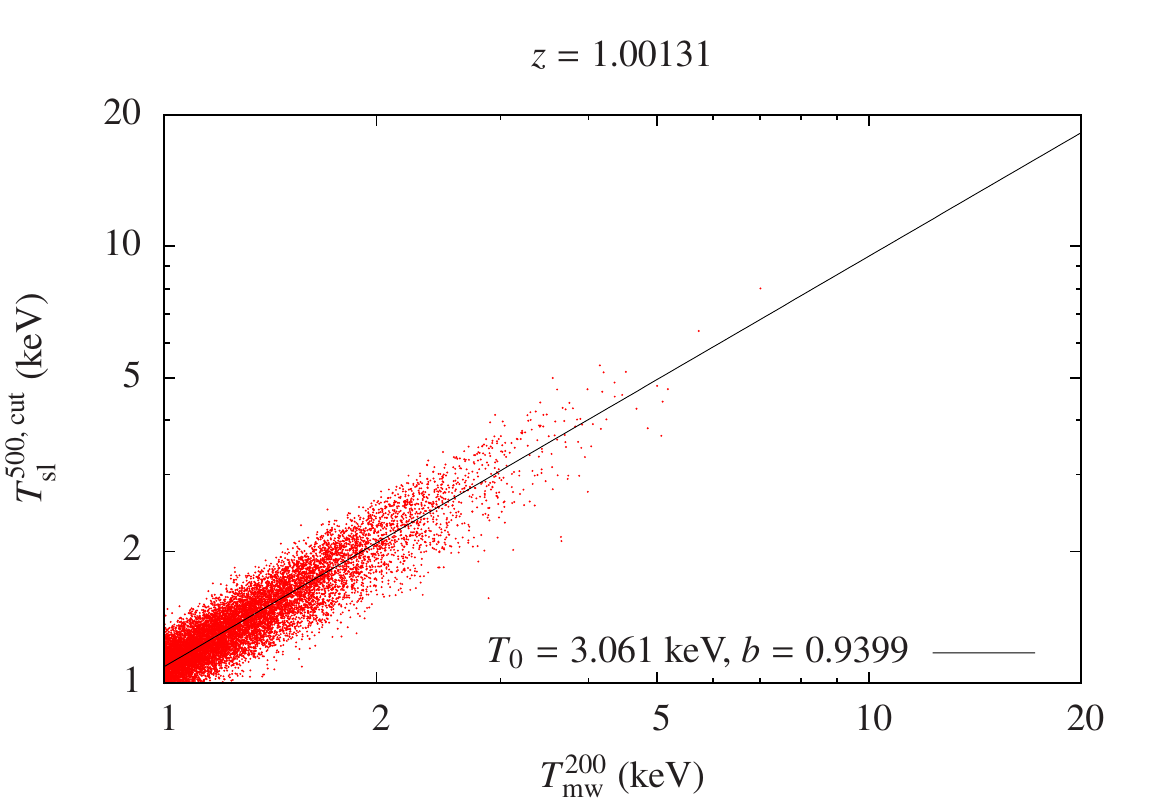}\\[2mm]
\includegraphics[width=0.33\textwidth]{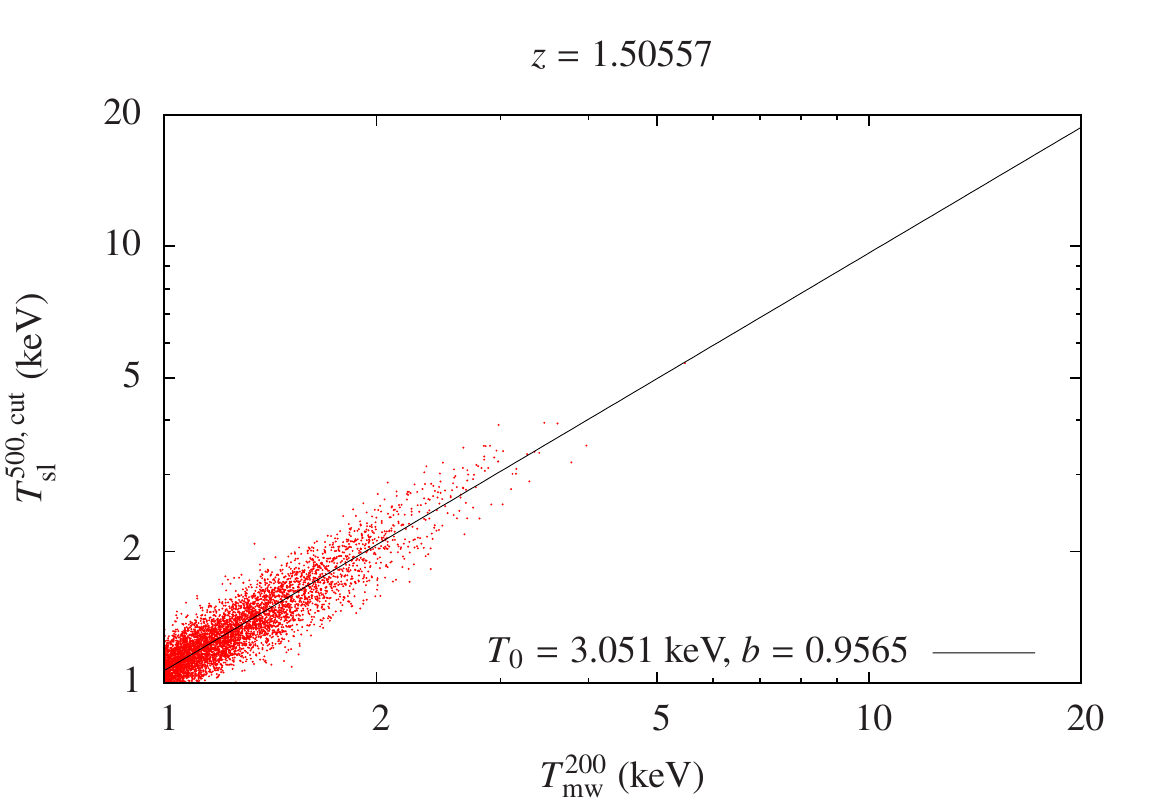}~\includegraphics[width=0.33\textwidth]{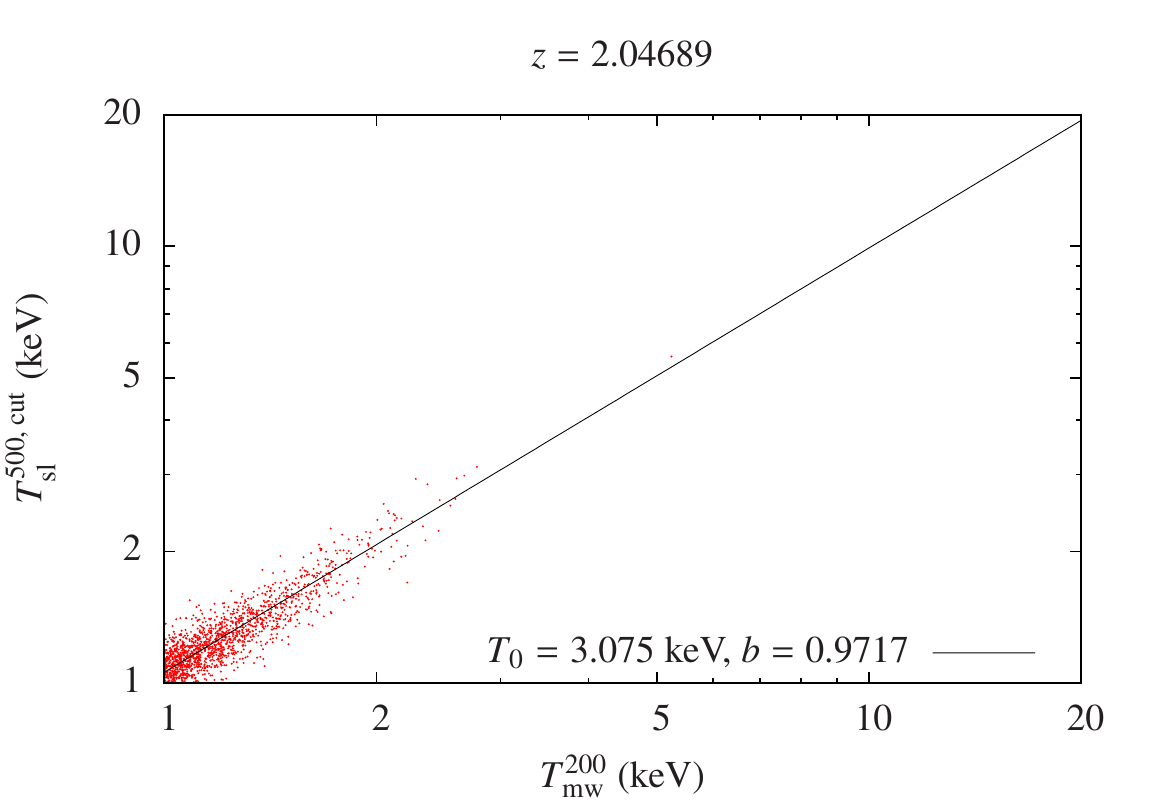}
\caption{Relation between $T_\rmn{mw}^{200}$ and $T_\rmn{sl}^\rmn{500,\,cut}$ inferred from the halo catalogue 
including fits of the form \eqref{eq:Trelation} for five different redshifts.}
\label{fig:scaledT}
\end{figure*}

To relate the temperature that is measured by observers ($\sim T_\rmn{sl}^\rmn{500,\,cut}$) to the one that our model is based on ($\sim T_\rmn{mw}^{200}$), we fitted a relation for each of the five different redshifts of the form
\begin{equation}
\label{eq:Trelation}
T_\rmn{sl}^\rmn{500,\,cut}=T_0\left(\frac{T_\rmn{mw}^{200}}{3\ \text{keV}}\right)^b,
\end{equation}
based on the cluster catalogue extracted from our simulation. The fitting parameters $T_0$ and $b$ therefore vary 
with redshift. We used only those clusters for the fit that fulfil both $T>1$~keV and 
$M_\rmn{DM}>1.17\times10^{13}\ h^{-1}\ \rmn{M}_\odot$ for both $r\in[0.15,1]R_{500}$ and $r<R_{200}$, where 
$M_\rmn{DM}$ is the mass of the cluster's DM component. The cut in mass corresponds to the requirement of 
having a halo that consists of at least 100 particles. The fitted functions \eqref{eq:Trelation} together 
with the data points the fits are based on are shown for various redshifts in Fig.~\ref{fig:scaledT}.

The resulting fitting parameters $T_0$ and $b$ can be found in Table~\ref{tab:fit}. Except for $z=0$, the parameter 
$T_0$ is very close to 3 keV. Additionally, the higher the redshift, the closer to unity is the parameter $b$. The rms of the deviations of individual clusters from the fitting relation \eqref{eq:Trelation} is with 0.21~keV largest for $z=0$ and with 0.12~keV smallest for $z=2$ and hence relatively small.

\begin{table}
\caption{The best-fitting parameters $T_0$ and $b$ as well as their errors $\Delta T_0$ and $\Delta b$, respectively, as a 
function of redshift $z$.}
\label{tab:fit}
\begin{center}
\begin{tabular}{ccccc}
\hline
$z$       & $T_0$ (keV) & $b$    & $\Delta T_0$ (keV)& $\Delta b$\\ \hline
0         & 3.209       & 0.9012 & 0.003             & 0.0013\\
0.507308  & 3.085       & 0.9114 & 0.005             & 0.0020\\
1.00131   & 3.061       & 0.9399 & 0.007             & 0.0029\\
1.50557   & 3.051       & 0.9565 & 0.012             & 0.0048\\
2.04689   & 3.075       & 0.9717 & 0.023             & 0.0087\\ \hline
\end{tabular}
\end{center}
\end{table}

\begin{figure*}
\centering
\includegraphics[width=0.33\textwidth]{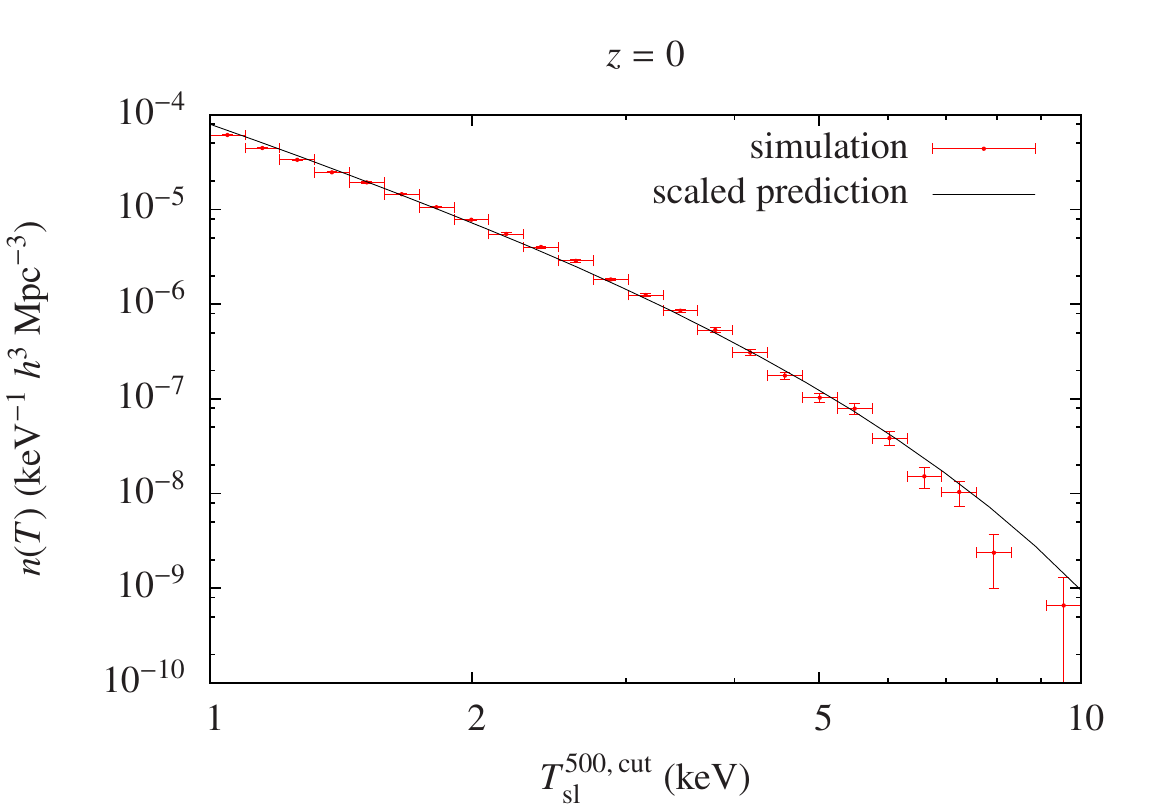}~\includegraphics[width=0.33\textwidth]{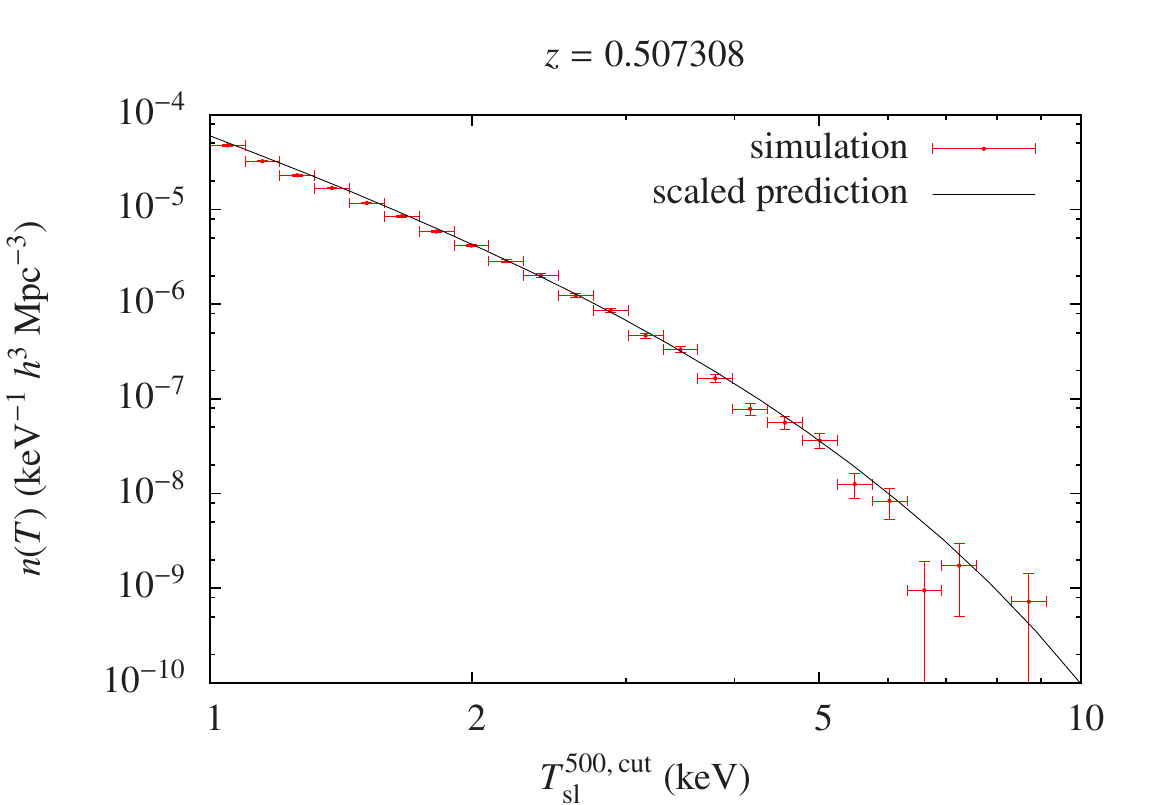}~
\includegraphics[width=0.33\textwidth]{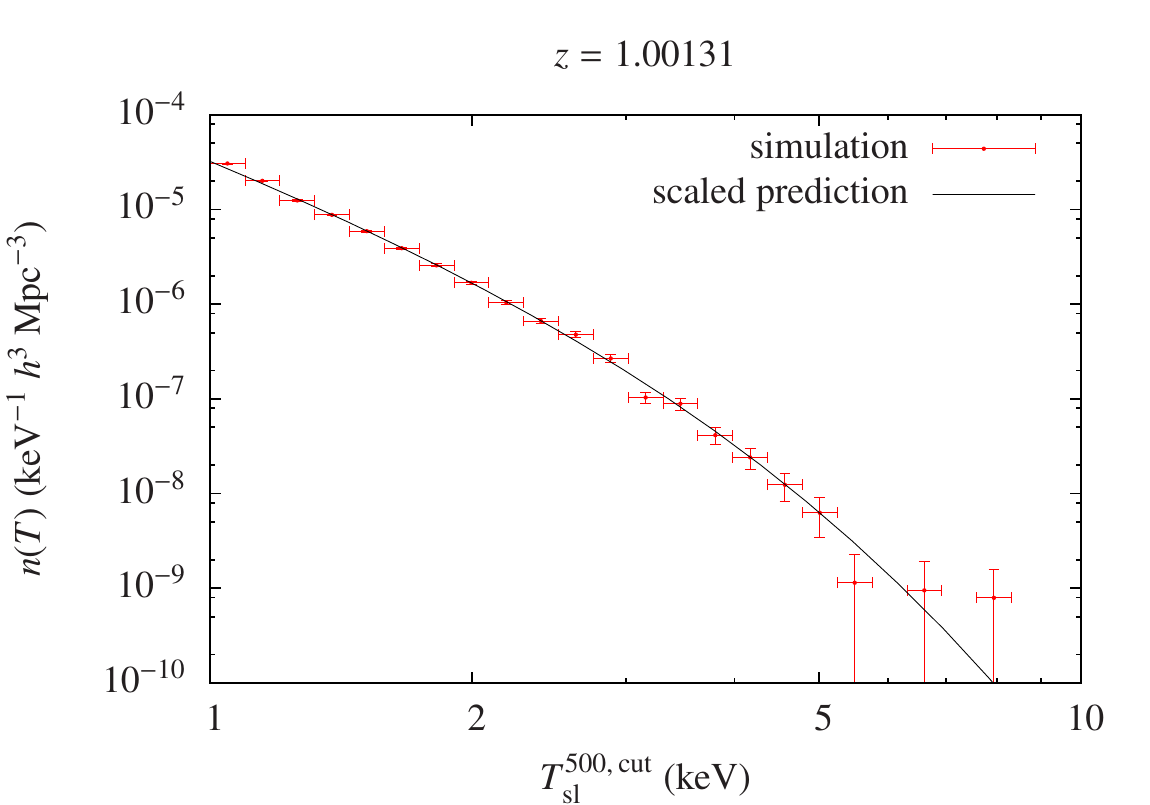}\\[2mm]
\includegraphics[width=0.33\textwidth]{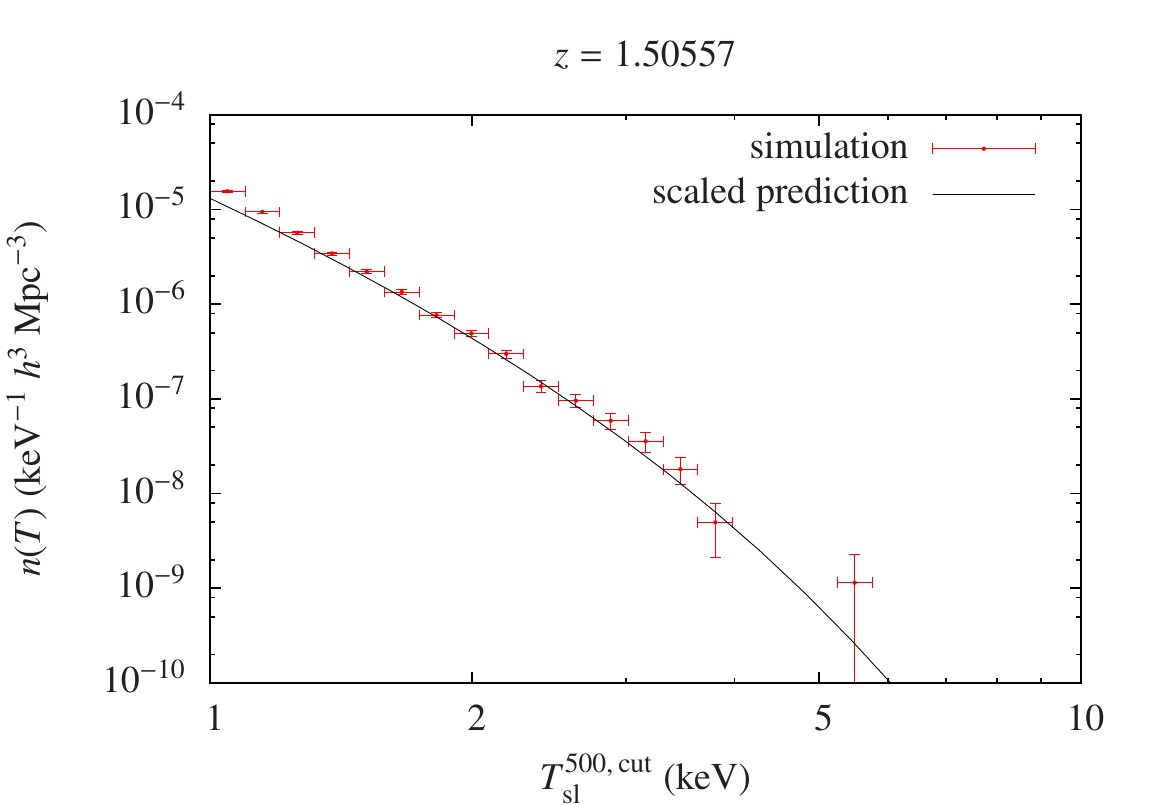}~\includegraphics[width=0.33\textwidth]{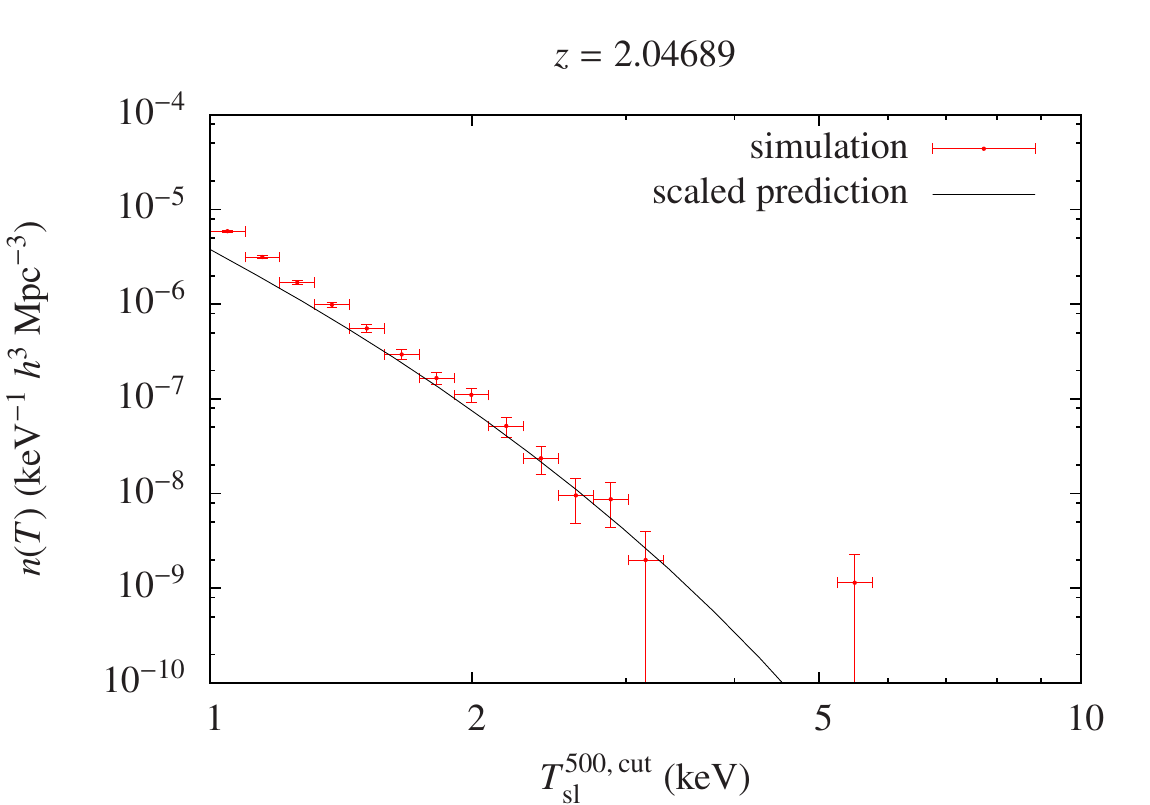}
\caption{Same as Fig.~\ref{fig:Tsl500cut}, but the theoretical temperature function was rescaled according to the redshift-dependent
relation between $T_\rmn{mw}^{200}$ and $T_\rmn{sl}^\rmn{500,\,cut}$ inferred from Fig.~\ref{fig:scaledT}.}
\label{fig:Tsl500cutRescaled}
\end{figure*}

Based on the relation \eqref{eq:Trelation}, we recalculated our theoretical prediction by scaling the temperature 
function accordingly, i.\,e.\ given $T_\rmn{sl}^\rmn{500,\,cut}$, we calculated $T_\rmn{mw}^{200}$ and used this 
temperature as input for equation~\eqref{eq:numDensT}. In doing so, an additional factor 
$|\dd T_\rmn{mw}^{200}/\dd T_\rmn{sl}^\rmn{500,\,cut}|$ had to be taken into account. The comparison between the 
rescaled theoretical temperature 
prediction and the one inferred from the simulation based on $T_\rmn{sl}^\rmn{500,\,cut}$ can be found in Fig.~\ref{fig:Tsl500cutRescaled}.

After a temperature rescaling in the theoretical prediction for the X-ray temperature function, the agreement
between the theoretical predictions and the numerical results based on $T_\rmn{sl}^\rmn{500,\,cut}$ is much better for 
all redshifts. Thus, by taking the redshift-dependent temperature conversion (equation~\ref{eq:Trelation}) into account, our 
theoretical model based on the statistics of gravitational-potential perturbations is consistent with numerical results 
that mimic as well as possible the X-ray temperature function inferred by real observations.

\section{Constraining cosmological parameters}
\label{sec:constraints}

Based on the results from the previous section, we want to determine the cosmological parameters $\sigma_8$ and 
$\Omega_\rmn{m}$ from a cluster sample by \citet{Vikhlinin2009}, which was also considered in the \textit{Planck} 2013 analysis of the Sunyaev-Zel'dovich (SZ) effect \citep{PlanckSZ2013}, using our theoretical model for the X-ray temperature function together with a proper temperature scaling dependent on redshift based on equation~\eqref{eq:Trelation}. The 
statistical analysis is very similar to the one already presented in \citet{Angrick2012}.

This section is meant as a proof of concept. To reliably constrain cosmological parameters in the future, one should stick to empirical relations between the potential of a cluster and its temperature, e.g.\ by combining data from gravitational lensing and X-ray temperature measurements, and not rely on relations inferred only from hydrodynamical simulations. See also the end of Sect.~\ref{sec:summary} for more details.

\subsection{The sample}
\label{subsec:sample}

The sample by \citet{Vikhlinin2009} consists of two subsamples, one at high and one at low redshift, based on 
\textit{ROSAT} PSPC All-Sky (RASS) and 400 deg$^2$ data. The \emph{low-redshift sample} is based on archival \textit{Chandra} data and consists of 49 clusters with 
flux $f>1.3\times10^{-11}$ erg s$^{-1}$ cm$^{-2}$ in the 0.5--2~keV band from several samples of RASS with a total 
area of 8.14 sr (26722 deg$^2$). The redshift coverage is $0.025<z<0.25$ with $\langle z\rangle\approx0.05$, and temperatures are in the range 
$2.61\ \text{keV}<T<14.72\ \text{keV}$ as inferred from their spectra measured in multiple annuli with \textit{Chandra}.

The \emph{high-redshift sample} consists of 36 clusters from the \textit{ROSAT} 400 deg$^2$ survey \citep{Burenin2007} further analysed with \textit{Chandra}
in the redshift range $0.35<z<0.9$ with $\langle z\rangle\approx0.5$ and a redshift-dependent flux limit in the 0.5--2 
keV band. For $z>0.473$, the limiting flux is $1.4\times10^{-13}$ erg s$^{-1}$ cm$^{-2}$, whereas for $z<0.473$, the 
flux limit corresponds to a minimal X-ray luminosity of $L_\rmn{X,min}=4.8\times10^{43}(1+z)^{1.8}$ erg s$^{-1}$. The 
temperatures of the clusters are in the range $2.13\ \text{keV}<T<11.08\ \text{keV}$ as inferred from their spectra measured with \textit{Chandra} in the region $[0.15,1]R_{500}$, where $R_{500}$ is the radius that encloses a mean overdensity of 500 times the critical density of the Universe.

The effective differential search volume $\dd V/\dd z$ as a function of mass $M$ and cosmological parameters 
$\Omega_\rmn{m}$, $\Omega_\Lambda$ and $H_0=72$ km s$^{-1}$ Mpc$^{-1}$ for both subsamples was made available in 
electronic form on a grid by A.~Vikhlinin. To convert it to a function of temperature, we used the best-fitting values of 
the mass--temperature relation of \citet{Vikhlinin2009},
\begin{equation}
\label{eq:MT}
M_{500}=M_0\left(\frac{T}{5~\text{keV}}\right)^\alpha E^{-1}(z),
\end{equation}
where $M_0=(3.02\pm0.11)\times10^{14}\ h^{-1}\ \rmn{M}_\odot$ and $\alpha=1.53\pm0.08$.

\subsection{The fitting procedure}
\label{subsec:fit}

Since the errors on the cluster number counts are Poissonian, we used the \emph{$C$ statistic} of \citet{Cash1979} 
for unbinned data to find the best-fitting values for $\Omega_\rmn{m}$ and $\sigma_8$, assuming a spatially flat universe, 
hence $\Omega_\Lambda=1-\Omega_\rmn{m}$. In the next section, we compare with the results from the 2015 data release of 
the Planck Collaboration so that we set the baryon density parameter $\Omega_\rmn{b}\,h^2=0.02225$, the index 
of the primordial power spectrum $n_\rmn{s}=0.9645$ and $H_0=67.27\ \text{km}\ \text{s}^{-1}\ \text{Mpc}^{-1}$ as 
inferred from the TT,TE,EE+lowP data \citep{Planck2015}. To include baryonic effects, we used the 
transfer function by \citet{Eisenstein1998} when computing the power spectrum $P_\delta(k)$.

The $C$ statistic is defined as
\begin{equation}
 \label{eq:defC}
C\equiv2\left(N-\sum_i\ln n_i\right),
\end{equation}
where $N$ is the total number of objects expected from the sample assuming a theoretical model, and $n_i$ is the 
theoretically expected differential number density of the $i$th cluster in the sample with temperature $T_i$ and 
redshift $z_i$. The sum extends over all sample members.

Although the cosmological parameters in the further discussion will slightly differ from their numerical values in our cosmological simulation, we assume that the temperature relation \eqref{eq:Trelation} does not change significantly so that we could still use the best-fitting values from Table \ref{tab:fit}. To take the listed uncertainties into account, we convolved with a normal distribution of 
the form
\begin{equation}
 \label{eq:normalTT}
p(T|T_\rmn{mw}^{200})\,\dd T=\frac{1}{\sqrt{2\pi}\,\sigma_T}
\exp\left\{-\frac{\left[T-T_\rmn{sl}^\rmn{500,\,cut}(T_\rmn{mw}^{200})\right]^2}{2\sigma_T^2}\right\}\,\dd T,
\end{equation}
where $T_\rmn{sl}^\rmn{500,\,cut}(T_\rmn{mw}^{200})$ is given by equation~\eqref{eq:Trelation}. Both $T_0$ and $b$ in 
that relation depend on redshift as follows. Let $z_n$ and $z_{n+1}$ be two consecutive redshifts for which the former 
relation can be deduced from the simulation. For a redshift $z$ with $z_n\leq z\leq z_{n+1}$, the values of the 
parameters $T_0$ and $b$ as well as their errors are linearly interpolated from their values at $z_n$ and $z_{n+1}$ 
(cf.\ Table~\ref{tab:fit}). To properly take the errors on the parameters $T_0$ and $b$ into account, the standard 
deviation was set to 
\begin{equation}
\sigma_T=T_\rmn{sl}^\rmn{500,\,cut}(T_\rmn{mw}^{200})\sqrt{\left(\frac{\Delta T_0}{T_0}\right)^2+
\left(\Delta b\ln\frac{T_\rmn{mw}^{200}}{3\ \text{keV}}\right)^2}
\end{equation}
due to Gaussian error propagation. Thus, the expected number of objects in each subsample is given by
\begin{align}
\label{eq:N}
N_\rmn{low|high}&=\int_{z_1}^{z_2}\dd z\int_{T_1}^{T_2}\dd T\,\frac{\dd V_\rmn{low|high}}{\dd z}(T,z)\\ \nonumber
&\times\int\dd T_\rmn{mw}^{200}\, n(T_\rmn{mw}^{200})\,p(T|T_\rmn{mw}^{200}),
\end{align}
where the subscripts `low' and `high' denote the low- and the high-redshift subsample, respectively. The integral 
boundaries depend on the subsample and are given in Sect.~\ref{subsec:sample} for $z$ and $T$. The integration over 
$T_\rmn{mw}^{200}$ has to be done over the whole valid range of $p(T|T_\rmn{mw}^{200})$.

Finally, the expected differential number density of the $i$th cluster is simply given by the convolution
\begin{align}
\label{eq:ni}
n_{i,\rmn{low|high}}&=\frac{\dd V_\rmn{low|high}}{\dd z}(T_i,z_i) \\ \nonumber
&\times\int\dd T_\rmn{mw}^{200}\int\dd T\,n(T_\rmn{mw}^{200})\,p(T|T_\rmn{mw}^{200})\,q(T_i|T),
\end{align}
with
\begin{equation}
\label{eq:q}
q(T_i|T)=\frac{1}{\sqrt{2\pi}\,\sigma_i}\exp\left[-\frac{(T_i-T)^2}{2\sigma_i^2}\right],
\end{equation}
where $\sigma_i$ is the measurement error of the $i$th cluster.

To jointly fit both the low- and the high-redshift cluster samples of \citet{Vikhlinin2009}, we had to add the two 
contributions, resulting in
\begin{equation}
\label{eq:CVikhlinin}
C=2\left(N_\rmn{low}-\sum_i \ln n_{i,\rmn{low}}+N_\rmn{high}-\sum_j \ln n_{j,\rmn{high}}\right).
\end{equation}
We searched for minima of the $C$ statistic as a function of the two cosmological parameters $\Omega_\rmn{m}$ and 
$\sigma_8$, which enter both via $n(T_\rmn{mw}^{200})$ and the volume factor $\dd V/\dd z$.

\citet{Cash1979} showed that one can create confidence intervals for the $C$ statistic exactly in the same way as it can be 
done for a $\chi^2$ fit using properties of the $\chi^2$ distribution. Following the work by \citet*{Lampton1976}, intervals with 
confidence $y$ are implicitly given solving
\begin{equation}
 \label{eq:confidence}
y=\int_0^t\dd\chi^2\,f(\chi^2),
\end{equation}
for $t$, where $f$ is the density of the $\chi^2_p$ distribution with $p$ degrees of freedom determined by the number 
of parameters. For 68 per cent confidence and $p=2$, it follows that $t=2.3$, while for 95 per cent confidence, $t=5.991$. Using 
the minimum of the $C$ statistic, $C_\rmn{min}$, we could simply calculate the 68 and 95 per cent confidence contours 
by searching for points in the parameter space for which $C=C_\rmn{min}+2.3$ and $C=C_\rmn{min}+5.991$, respectively.

\section{Results}
\label{sec:results}

\begin{figure}
\centering
\includegraphics[width=84mm]{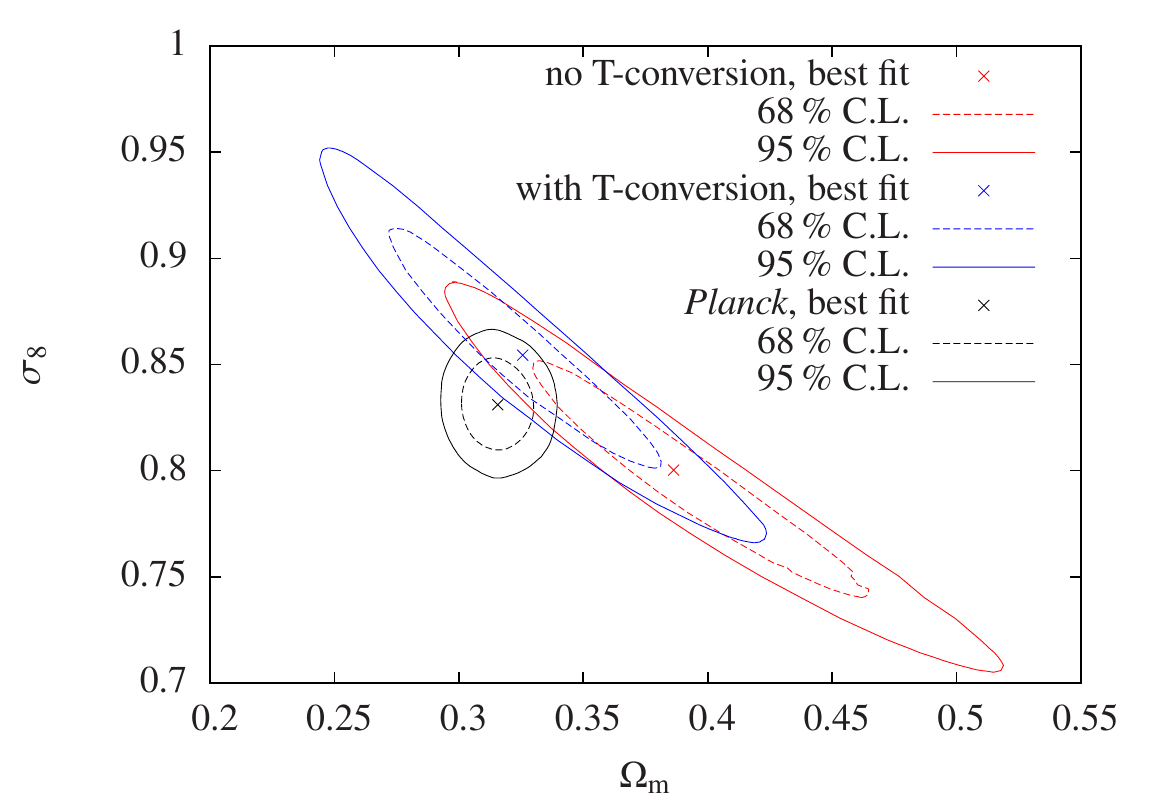}
\caption{68 and 95 per cent confidence contours for the parameters $\Omega_\rmn{m}$ and $\sigma_8$ based on our cluster 
temperature function as inferred from the two subsamples by \citet{Vikhlinin2009} excluding (red contours) and 
including (blue contours) the temperature conversion \eqref{eq:Trelation}. For comparison we also provide the purely CMB-based
\textit{Planck} TT,TE,EE+lowP results \citep[][black contours]{Planck2015}.}
\label{fig:contours}
\end{figure}

In Fig.~\ref{fig:contours}, we compare the 68 and 95 per cent confidence contours for the parameters $\Omega_\rmn{m}$ 
and $\sigma_8$ inferred with our semi-analytic cluster temperature function to the respective confidence contours from 
the \textit{Planck} 2015 data release \citep{Planck2015} based on the TT,TE,EE+lowP data both including the 
temperature relation \eqref{eq:Trelation} and excluding it by setting 
$p(T|T_\rmn{mw}^{200})=\updelta_\rmn{D}(T-T_\rmn{mw}^{200})$ 
in equations~\eqref{eq:N} and~\eqref{eq:ni} for the latter case.

Taking the temperature conversion from $T_\rmn{sl}^\rmn{500,\,cut}$ to $T_\rmn{mw}^{200}$ into account when fitting 
our theoretical X-ray temperature function to the X-ray sample by \citet{Vikhlinin2009} shifts the confidence contours 
to smaller values for $\Omega_\rmn{m}$ and to larger values for $\sigma_8$ compared to the case where such a 
temperature conversion based on the difference between measured and theoretically motivated temperature is neglected. 
Additionally, the sizes of both the 68 and the 95 per cent contour are smaller if the temperature conversion is 
included.

The \textit{Planck} constraints are in agreement with our results incorporating the redshift-dependent temperature 
conversion \eqref{eq:Trelation} at 1$\sigma$-level, whereas the results without the temperature conversion are in 
agreement with the \textit{Planck} results only at the 2$\sigma$-level. Hence, discriminating between the temperature that is actually
inferred from an X-ray measurement (similar to $T_\rmn{sl}^\rmn{500,\,cut}$) and theoretically motivated temperature 
consistent with the virial theorem (similar to $T_\rmn{mw}^{200}$) seems crucial when our potential-based X-ray 
temperature function is used to constrain cosmological parameters.

\section{Summary \& conclusions}
\label{sec:summary}

In the first part of this article, we further refined the X-ray temperature function for clusters based on the cosmic 
gravitational potential by \citet{Angrick2009,Angrick2012} by incorporating the redshift evolution of the temperature 
function for virialized structures (equation~\ref{eq:numDensT}) into our analytical merger model. We then compared our 
theoretical model to X-ray temperature functions inferred from a fully hydrodynamical simulation based on two different 
temperature definitions at five different redshifts: (1) the mass-based temperature inside 
$R_{200}$, $T_\rmn{mw}^{200}$, (2) the spectroscopic like temperature inside $R_{500}$ with the inner part 
$<0.15\,R_{500}$ cut out, $T_\rmn{sl}^\rmn{500,\,cut}$. The main results can be summarized as follows:
\begin{enumerate}
\item Our theoretical temperature function is in very good agreement with the numerical one based on 
$T_\rmn{mw}^{200}$ except for relatively high temperatures which is presumably due to resolution effects of the 
simulation in the high-mass regime.
\item Compared to the numerical temperature function based on $T_\rmn{sl}^\rmn{500,\,cut}$, our theoretical X-ray 
temperature function underestimates the abundance of objects over a large temperature range for all redshifts examined.
\item For each redshift $z$ separately it is possible to find a relatively tight relation of the form 
$T_\rmn{sl}^\rmn{500,\,cut}=T_0(T_\rmn{mw}^{200}/3\ \text{keV})^b$, where both $T_0$ and $b$ are fitting parameters 
depending on $z$. The rms of the deviations of individual clusters in the simulation from the above relation reaches from 0.12~keV at $z=2$ to 0.21~keV at $z=0$.
\item Scaling the temperature of our theoretical temperature function according to the above relation for each 
redshift and then comparing it to the numerical temperature function based on $T_\rmn{sl}^\rmn{500,\,cut}$ yields very 
good agreement between both functions.
\end{enumerate}
The second part of this article was meant as a proof of concept. Here, we used the redshift-dependent temperature scaling found in the first part for our 
theoretical model to fit it to an X-ray sample by \citet{Vikhlinin2009} and constrain the cosmological parameters 
$\Omega_\rmn{m}$ and $\sigma_8$. For comparison we also fitted these parameters without taking the aforementioned 
temperature conversion into account. The main results are the following:
\begin{enumerate}
\setcounter{enumi}{4}
\item Incorporating the temperature conversion shifts the confidence contours in the $\Omega_\rmn{m}$-$\sigma_8$ plane 
towards smaller values of $\Omega_\rmn{m}$ and larger values of $\sigma_8$. Additionally, the size of the contours is 
reduced.
\item We find agreement at 1$\sigma$-level between the purely CMB-based TT,TE,EE+lowP results from the \textit{Planck} 2015 data release \citep{Planck2015} and our theoretical model incorporating the temperature conversion, whereas the two only agree at 2$\sigma$-level with each other if the temperature conversion is neglected.
\end{enumerate}
Concluding, it seems necessary to establish an empirical relation between the potential and the X-ray temperature of a cluster since relying solely on a scaling inferred from a hydrodynamical simulation may not allow to constrain cosmological parameters reliably enough. Combining data from gravitational lensing and X-ray observations of both relaxed and merging galaxy clusters could be a promising way to find such a relation since gravitational lensing probes the gravitational potential of a cluster projected along the line-of-sight, whereas one can infer the temperature of its intracluster medium (ICM) from X-ray spectra.

Once such an empirical relation is established, we will use our potential-based temperature function on more recent and 
bigger samples of X-ray clusters to constrain $\Omega_\rmn{m}$ and $\sigma_8$ from cluster cosmology with higher 
accuracy and reliability. However, finding a well-defined physical and analytical model for this relation would be even more desirable 
since it would allow directly the modelling of the temperature abundance of galaxy clusters, thus avoiding additional sources of scatter.

\section*{Acknowledgements}
We thank Klaus Dolag for providing the initial conditions of the simulation and Margherita Grossi for the code to run 
them. Simulations and cluster catalogues were created on the Intel SCIAMA High Performance Compute (HPC) cluster 
which is supported by the ICG, SEPNet and the University of Portsmouth.

\bibliography{bibliography.bib}

\end{document}